\documentclass[
  10pt,  
  aps,  
  amsmath,
  amssymb,
  twocolumn,
  prb,
  superscriptaddress,
  floatfix
]{revtex4-1} 

\usepackage{graphicx}
\usepackage{mathtools}
\usepackage{natbib}
\usepackage{hyperref}

\newcommand{\be}{\begin{equation}}
\newcommand{\e}{\end{equation}}
\newcommand{\beml}{\begin{subequations}}
\newcommand{\eml}{\end{subequations}}
\newcommand{\beq}{\begin{eqnarray}}
\newcommand{\eq}{\end{eqnarray}}
\newcommand{\ba}{\begin{array}}
\newcommand{\ea}{\end{array}}
\newcommand{\bpm}{\begin{pmatrix}}
\newcommand{\epm}{\end{pmatrix}}
\newcommand{\bc}{\begin{cases}}
\newcommand{\ec}{\end{cases}}
\newcommand{\lt}{\left}
\newcommand{\rt}{\right}

\newcommand{\la}{\langle}
\newcommand{\ra}{\rangle}
\newcommand{\ep}{\varepsilon}

\newcommand{\bb}{\boldsymbol}

\catcode`@=11

\def\nvphantom{\v@true\h@false\nph@nt}
\def\nhphantom{\v@false\h@true\nph@nt}
\def\nphantom{\v@true\h@true\nph@nt}
\def\nph@nt{\ifmmode\def\next{\mathpalette\nmathph@nt}%
  \else\let\next\nmakeph@nt\fi\next}
\def\nmakeph@nt#1{\setbox\z@\hbox{#1}\nfinph@nt}
\def\nmathph@nt#1#2{\setbox\z@\hbox{$\m@th#1{#2}$}\nfinph@nt}
\def\nfinph@nt{\setbox\tw@\null
  \ifv@ \ht\tw@\ht\z@ \dp\tw@\dp\z@\fi
  \ifh@ \wd\tw@-\wd\z@\fi \box\tw@}

\DeclareMathOperator{\tr}{Tr}
\DeclareMathOperator{\mtr}{tr}
\DeclareMathOperator{\diag}{diag}
\DeclareMathOperator{\im}{Im}

\DeclareMathOperator{\sign}{sign}

\DeclareMathOperator{\Span}{span}

\begin{document}

\title{Anisotropy of spin-transfer torques and Gilbert damping induced by Rashba coupling}

\author{I.\,A.~Ado}
\affiliation{Radboud University, Institute for Molecules and Materials, NL-6525 AJ Nijmegen, The Netherlands}
\author{P.\,M.~Ostrovsky}
\affiliation{Max Planck Institute for Solid State Research, Heisenbergstr.\,1, 70569 Stuttgart, Germany}
\affiliation{L.\,D.~Landau Institute for Theoretical Physics RAS, 119334 Moscow, Russia}
\author{M.~Titov}
\affiliation{Radboud University, Institute for Molecules and Materials, NL-6525 AJ Nijmegen, The Netherlands}
\affiliation{ITMO University, Saint Petersburg 197101, Russia}

\begin{abstract}
Spin-transfer torques (STT), Gilbert damping (GD), and effective spin renormalization (ESR) are investigated microscopically in a 2D Rashba ferromagnet with spin-independent Gaussian white-noise disorder. Rashba spin-orbit coupling-induced anisotropy of these phenomena is thoroughly analysed. For the case of two partly filled spin subbands, a remarkable relation between the anisotropic STT, GD, and ESR is established. In the absence of magnetic field and other torques on magnetization, this relation corresponds to a current-induced motion of a magnetic texture with the classical drift velocity of conduction electrons. Finally, we compute spin susceptibility of the system and generalize the notion of spin-polarized current.
\end{abstract}

\maketitle

Possibility to efficiently manipulate magnetic order by means of electric current has gained a lot of attention over the past decades~\cite{Brataas2012review, Jungwirth2016AFMreview}. Potential applications include race track memory~\cite{Parkin-race-track-2008, Parkin-race-track-2015}, spin torque magnetization switching~\cite{Miron2011FMswitching, Wadley2016AFMswitching}, skyrmion-based technology~\cite{Kiselev2011, Fert2013skyrmion-racetrack}, and other promising concepts. Spintronic logic and memory devices based on current-driven magnetization dynamics are believed to achieve high speed, low volatility, outstanding durability, and low material costs with promises to outperform charge-trapping solid-state memory devices\cite{Chappert2007review}.

In the light of recent detection of fast domain wall (DW) motion in magnetic films~\cite{Miron2011-400m/s, Parkin2015-750m/s} and predictions of even higher DW velocities in antiferromagnets~\cite{Gomonay2016AFM}, current-induced dynamics of domain walls, skyrmions, and other magnetic textures remain an important research subject in the field of spintronics. Such dynamics is mainly determined by the interplay of the two phenomena: Gilbert damping (GD) and spin torques~\cite{DuinevsKeldysh2007, alphavsbetaEXP2008, Tserkovnyak2008review, Tatara2008review}.

In the absence of spin-orbit coupling (SOC), spin torques emerge only in the systems with nonuniform magnetization profiles and are most often referred to as spin-transfer torques (STT). At the same time, the classification of spin torques usually gets more complicated if coupling between spin and orbital degrees of freedom becomes pronounced. Moreover, the debate on the microscopic origin of spin torques in the latter case remains ongoing~\cite{Sinova2015review, Manchon2019review}. Below, we regard STT, in the continuum limit, as a contribution to the total torque on magnetization that is linear with respect to both the electric field~$\bb E$ and the first spatial derivatives of the unit vector of magnetization direction $\bb n$. We note that, in the absence of SOC, physics of STT is well understood~\cite{Tserkovnyak2008review, Tatara2008review}.

In a similar fashion, Gilbert damping may be generally associated with the terms of the Landau-Lifshitz-Gilbert (LLG) equation that are odd under time reversal and linear with respect to the time derivative of $\bb n$. In the most simplistic approach, GD is modeled by a single phenomenological term $\alpha\bb n \times \partial_t \bb n$ that corresponds to ``isotropic'' damping.

However, it has been known for quite a while that GD may exhibit anisotropic behaviour~\cite{Safonov2002, Meckenstock2004GDanisotropy, Steiauf2005anisotropicGDabinitio, Gilmore2010anisotropicGDabinitio, Mankovsky2013, Hals2014anisotropy, Kasatani2014GDanisotropy, Kasatani2015GDanisotropy, ManchonGD2017}. Or, to be more precise, that the scalar damping constant $\alpha$, in general, should be replaced by a damping matrix with the components depending on the orientation of $\bb n$. These two manifestations of anisotropy may be referred to as rotational and orientational anisotropy, respectively~\cite{Gilmore2010anisotropicGDabinitio}. Experimental observation of the orientational anisotropy of Gilbert damping has been reported very recently in a metal ferromagnet (FM)/semiconductor interface of Fe/GaAs(001)~\cite{Chen2018} and in epitaxial CoFe films~\cite{Li2019}. The authors of Ref.~[\onlinecite{Chen2018}] argued that the measured anisotropy rooted in the interplay of interfacial Rashba and Dresselhaus spin-orbit interaction.

Given the equal importance of GD and STT in the context of current-induced magnetization dynamics and the significant progress made in the understanding of the anisotropic nature of Gilbert damping, we find it surprising that the anisotropy of spin-transfer torques has so far only been addressed phenomenologically~\cite{Hals2013anisotropy, Hals2014anisotropy}.

In the present paper, we consider a 2D Rashba FM with spin-independent electron scattering. A microscopic analysis, performed for an arbitrary magnetization direction, allows us to quantify the rotational as well as the orientational anisotropy of both STT and GD induced by Rashba SOC. Our results indicate that, for a Rashba FM system, spin-transfer torques~$\bb T^{\text{STT}}$ and Gilbert damping~$\bb T^{\text{GD}}$ entering the LLG equation
\be
\label{LLG_short}
\partial_t\bb n=\gamma \bb n \times \bb{H}_{\text{eff}}+\bb T^{\text{STT}}+\bb T^{\text{GD}}+\dots
\e
naturally acquire the following forms:
\begin{subequations}
\label{STT_and_GD_general}
\begin{align}
\label{STT_general}
\bb T^{\text{STT}}&=
\xi_0 \partial_{\bb v}\bb n
-\xi_{\parallel}[\bb n\times \partial_{\bb v}\bb{n}_{\parallel}]-\xi_{\perp}[\bb n \times \partial_{\bb v}\bb{n}_{\perp}],
\\
\label{GD_general}
\bb T^{\text{GD}}&=
\xi_0 \partial_{\,t \phantom{\bb v}\nphantom{\,t}}\bb n
-\xi_{\parallel}[\bb n\times \partial_{\,t \phantom{\bb v}\nphantom{\,t}}\bb n_{\parallel}]
-\xi_{\perp}[\bb n \times \partial_{\,t \phantom{\bb v}\nphantom{\,t}}\bb n_{\perp}],
\end{align}
\end{subequations}
where $\xi_{i}=\xi_{i}(\bb n)$, the operator $\partial_{\bb v}=(\bb v_{\text{d}}\cdot\bb\nabla)$ is expressed via the classical electron drift velocity $\bb v_{\text{d}}=e\bb E\hbar\tau/m$, and $\bb n_{\parallel/\perp}$ stands for the in-plane/perpendicular-to-the-plane component of the vector field $\bb n$:
\be
\label{components_of_n}
\bb n=\bb n_{\parallel}+\bb n_{\perp}, \qquad \bb n_{\perp}=\bb e_z n_z=\bb e_z \cos{\theta}.
\e
For convenience, we have included the term $\xi_0\partial_t\bb n$ into the definition of~$\bb T^{\text{GD}}$. This term, being even under time reversal, leads to a renormalization of spin in the LLG equation~\cite{Tatara2008review} and does not contribute to damping. In what follows, we refer to such renormalization as effective spin renormalization (ESR).

The rotational and orientational anisotropy arising in Eqs.~(\ref{STT_and_GD_general}) appear to be a natural consequence of the fact that the Rashba spin-orbit interaction singles out the direction perpendicular to the electron 2D plane. The orientational anisotropy of the dimensionless functions $\xi_i(\bb n)$ is determined by all space symmetries of the system and, for a general Rashba FM, may turn out to be rather complex. However, for the particular interface model of the $C_{\infty v}$ symmetry class, which we consider below, one simply finds $\xi_i=\xi_i(n_z^2)$.

Before we proceed, let us describe at least two important outcomes of Eqs.~(\ref{STT_and_GD_general}). First, according to the usual convention, STT consist of two contributions: the adiabatic torque $\propto(\bb j_s\cdot\bb\nabla)\bb n$ and the nonadiabatic torque \mbox{$\propto\bb n\times(\bb j_s\cdot\bb\nabla)\bb n$}, where $\bb j_s$ denotes a spin-polarized current. For vanishing SOC, the adiabatic torque has a clear physical meaning. As far as spins of conduction electrons adiabatically follow local magnetization direction, the corresponding change of their angular momentum is transferred to the magnetic texture. Since $\uparrow$ and $\downarrow$ spins point in the opposite directions along $\bb n$, the transfer rate is proportional to $(\bb j_s\cdot\bb\nabla)\bb n$, where $\bb j_s=\bb j_\uparrow - \bb j_\downarrow$. In the presence of SOC, however, conduction spins are no longer aligned with the direction of $\bb n$ and, thus, the entire concept of spin-polarized current becomes somewhat vague. For the particular Rashba model, our results reveal an important relation between the adiabatic torque and ESR, providing steps toward better understanding of the former for systems with SOC.

Another remarkable property of Eqs.~(\ref{STT_and_GD_general}) is a simple and exact relation between the nonadiabatic torque and GD, which has an important implication for current-induced motion of magnetic textures (e.\,g., domain walls or skyrmions). Indeed, by transforming Eq.~(\ref{LLG_short}) into the moving reference frame~\cite{reference_frame_comment} $\bb r'=\bb r-\bb v_d t$, one immediately observes that both components of the nonadiabatic torque are exactly cancelled by the corresponding Gilbert damping terms. Therefore, if the effect of other driving torques on the motion of a magnetic texture is negligible, then its terminal velocity, in the moving reference frame, shall vanish for mediate currents~\cite{1stLiZhang2004, Thiaville2005} (in the absence of magnetic field). This implies that, in the laboratory reference frame, the texture moves with the universal electron drift velocity $\bb v_d$. Certainly, in the presence of, e.g., spin-orbit torques, which can assist motion of domain walls and skyrmions~\cite{Miron2011-400m/s, Kim2012}, the resulting dynamics might differ. In any case, the analysis of such dynamics can still be performed in the moving reference frame, where the effect of the nonadiabatic spin-transfer torque is conveniently absent. 

Having outlined our main results, we skip further discussion until Sec~\ref{sec::discussion}. The rest of the paper is organized as follows. In Sec.~\ref{sec::model} we introduce the model and use an expansion in spatial gradients to reduce the analysis to a study of a homogeneous system. Self-energy and Kubo formulas are addressed in Sec.~\ref{sec::disorder}. A~general relation between STT, GD, and ESR (in the considered model) is obtained in Sec.~\ref{sec::relation}, while in Sec.~\ref{sec::vector_forms} we establish the exact vector structures of these quantities. Some analytical insight into our general results is provided in Sec.~\ref{sec::closed_forms}~and~Sec.~\ref{sec::asymptotics}. An extensive Discussion of Sec~\ref{sec::discussion} is followed by Conclusions (and seven Appendices).

\section{Model}
\label{sec::model}
\subsection{Generalized torque in $s$-$d$ model}
In what follows, we adopt the ideology of the $s$-$d$ model by performing a decomposition of a FM into a system of localized spins $\bb S_i$ and a system of noninteracting conduction electrons. Despite being rather simplistic, this approach has proven to describe very well the key properties of current-induced magnetization dynamics in ferromagnetic systems~\cite{2ndLiZhang2004, Stiles2013DMIvsA, 2014Kurebayashi, 2014MokrousovSOT}.

If the value of $\vert\bb S_i\vert=S$ can be assumed sufficiently large, then it is natural to treat the localized spins classically by means of the unit vector $\bb n(\bb r_i)=\bb S_i/S$, which points in the opposite to local magnetization direction. In this case, the $s$-$d$-like local exchange interaction between the localized spins and conduction electrons is given, in the continuum limit, by
\be
\label{H_sd}
\mathcal{H}_{\text{sd}}=J_{\text{sd}} S\,\bb n(\bb r,t)\cdot\bb{\sigma},
\e
with $J_{\text{sd}}$ quantifying the strength of the exchange and Pauli matrices $\bb \sigma$ representing the spins of conduction electrons.

It is known~\cite{Tatara2008review} that interaction of the form of Eq.~(\ref{H_sd}), leads to the following LLG equation for the dynamics of the vector $\bb n$:
\be
\label{LLG_micro}
\partial_t\bb n=\gamma \bb n \times \bb{H}_{\text{eff}}+\frac{J_{\text{sd}}A}{\hbar}[\bb s(\bb r,t) \times\bb n(\bb r,t)],
\e
where $\gamma$ is the bare gyromagnetic ratio, $\bb{H}_{\text{eff}}$ describes the effective magnetic field, $A$ denotes the area of the magnet unit cell, and $\bb s(\bb r,t)$ stands for the nonequilibrium spin density of conduction electrons~\cite{comment_on_hbar/2}. The second term on the right hand side of Eq.~(\ref{LLG_micro}) represents the generalized torque on magnetization
\be
\label{torque_gen}
\bb T=\frac{J_{\text{sd}}A}{\hbar}[\bb s(\bb r,t) \times\bb n(\bb r,t)].
\e

Assuming slow dynamics of $\bb n(\bb r,t)$ on the scale of electron scattering time and smoothness of magnetization profile on the scale of electron mean free path, one may expand the generalized torque in time and space gradients of $\bb n$. In this paper, we consider two particular terms of such expansion,
\be
\label{torque_dec}
\bb T=\bb T^{\text{STT}}+\bb T^{\text{GD}}+\dots,
\e
ignoring all other contributions (such as, e.g., spin-orbit torques). In Eq.~(\ref{torque_dec}) and below, we identify spin-transfer torques $\bb T^{\text{STT}}$ as a double response of $\bb T$ to the electric field $\bb E$ and to the spatial gradients of $\bb n$, while the Gilbert damping vector $\bb T^{\text{GD}}$ (which also includes the ESR term) is defined as a response to the time derivative of $\bb n$,
\begin{subequations}
\label{STT_and_GD_def}
\begin{align}
\label{STT_def}
T^{\text{STT}}_\alpha &=\sum_{\beta\gamma\delta}{\mathcal T^{\text{STT}}_{\alpha\beta\gamma\delta}\, E_\beta \nabla_\gamma n_\delta},
\\
\label{GD_def}
T^{\text{GD}}_\alpha &=\sum_{\delta}{\mathcal T^{\text{GD}}_{\alpha\delta}\, \partial_t n_\delta}.
\end{align}
\end{subequations}
Microscopic analysis of the tensors $\mathcal T^{\text{STT}}$ and $\mathcal T^{\text{GD}}$ is the main subject of the present work. 


\subsection{Single particle problem}
\label{sec::SPproblem}
According to Eqs.~(\ref{STT_and_GD_def}), the vectors $\bb T^{\text{STT}}$ and $\bb T^{\text{GD}}$ represent linear response to the time derivative of magnetization direction and to the time derivative of vector potential, respectively. Hence, computation of both vectors can be performed with the help of Kubo formulas that make use of Green's functions of the corresponding time-independent problem. We choose the latter to originate in the 2D Rashba model~\cite{1984BychkovRashba} with the effective $s$-$d$-type term of Eq.~(\ref{H_sd}),
\be
\label{Hamiltonian}
\mathcal{H}=p^2/2m+\alpha_{\text{\tiny R}}\, [\bb{p}\times\bb{\sigma}]_z  + J_{\text{sd}} S\,\bb n(\bb r)\cdot\bb{\sigma},
\e
where $\alpha_{\text{\tiny R}}$ characterizes the strength of Rashba coupling and $m$ is the effective electron mass.

The Hamiltonian of Eq.~(\ref{Hamiltonian}) should be supplemented with a momentum relaxation mechanism since both STT and GD tensors, similarly to the conductivity tensor, contain essentially dissipative components. We assume that momentum relaxation in the system is provided by scattering on a spin-independent Gaussian white-noise disorder potential $V_{\text{dis}}(\bb r)$. Thus, the full Hamiltonian of a single conduction electron reads
\be
\mathcal{H}_{\text{dis}}=\mathcal{H}+V_{\text{dis}}(\bb r),
\e
where the disorder potential is characterized by the zero average $\la V_{\text{dis}}(\bb r)\ra=0$ and the pair correlator
\be
\label{V_correlator}
\lt\la V_{\text{dis}}(\bb{r}) V_{\text{dis}}(\bb{r}')\rt\ra = (\hbar^2/m\tau)\, \delta(\bb{r}-\bb{r}').
\e
The angular brackets in Eq.~(\ref{V_correlator}) stand for the averaging over the disorder realizations, $\tau$ is the mean scattering time measured in the inverse energy units.

One can readily observe from Eq.~(\ref{torque_gen}) that the generalized torque $\bb T$ can be understood as a spatial density of a disorder-averaged mean value of the operator $(J_{\text{sd}}A/\hbar)\hat{\bb T}$, where we refer to
\be
\label{T_operator}
\hat{\bb T} =\bb \sigma\times\bb n(\bb{r}),
\e
as the dimensionless torque operator.

\subsection{Expansion in spatial gradients}
Computation of STT involves the expansion of the Hamiltonian $\mathcal H$ of Eq.~(\ref{Hamiltonian}) and the corresponding Green's function
\be
\label{Gmathcal}
\mathcal G^{R,A}=(\varepsilon-\mathcal H\pm i 0)^{-1}
\e
in the first spatial gradients of $\bb n$ up to the linear terms. We obtain the latter utilizing the Taylor expansion
\be
\label{Taylor}
\bb n(\bb r)=\bb n(\bb r_\ast)+\sum\limits_{\gamma}(\bb r-\bb r_\ast)_\gamma\nabla_\gamma \bb n(\bb r_\ast),
\e
at some particular point $\bb r_\ast$.

With the help of Eq.~(\ref{Taylor}), $\mathcal H$ can be, then, approximated as
\be
\label{H_gradient}
\mathcal{H}=H+J_{\text{sd}} S\sum\limits_{\gamma}(\bb r-\bb r_\ast)_\gamma\nabla_\gamma \bb n(\bb r_\ast)\cdot\bb{\sigma},
\e
where the Hamiltonian
\be
\label{Hamiltonian_homo}
H=p^2/2m+\alpha_{\text{\tiny R}}\, [\bb{p}\times\bb{\sigma}]_z  +J_{\text{sd}} S\,\bb n(\bb r_\ast)\cdot\bb{\sigma}
\e
describes the homogeneous electronic system with a fixed direction of magnetization set by $\bb n(\bb r_\ast)$.

Similarly, we approximate the Green's function $\mathcal G^{R,A}$, employing the Dyson series
\begin{multline}
\label{G_gradient}
\mathcal G^{R,A}(\bb r,\bb r')=G^{R,A}(\bb r-\bb r')+
J_{\text{sd}} S\int d^2 r''\,
G^{R,A}(\bb r-\bb r'')\\\times
\Bigl[
\sum\limits_{\gamma}(\bb r''-\bb r_\ast)_\gamma\nabla_\gamma \bb n(\bb r_\ast)\cdot\bb{\sigma}
\Bigr]
G^{R,A}(\bb r''-\bb r')
\end{multline}
and the Green's function
\be
\label{G}
G^{R,A}=(\varepsilon-H\pm i 0)^{-1}
\e
that corresponds to the homogeneous system. Note that, in Eq.~(\ref{G_gradient}), we kept only the terms that are linear in the gradients of $\bb n$, as prescribed.

\subsection{Spectrum of the homogeneous system}
The spectrum of $H$ incorporates two spectral branches
\be
\label{spectrum}
\varepsilon_{\pm}(\bb p)=p^2/2m\pm\sqrt{\Delta_{\text{sd}}^2+(\alpha_{\text{\tiny R}}p)^2-2\varsigma\alpha_{\text{\tiny R}}\Delta_{\text{sd}} \,p\sin{\theta}\sin{\varphi}},
\e
where the angle $\theta$ stands for the polar angle of $\bb n$ with respect to the $z$ axis [see also Eq.~(\ref{components_of_n})], while $\varphi$ is the angle between the momentum $\bb p$ and the in-plane component of the vector~$\bb n$: $\varphi=\phi_{\bb p}-\phi_{\bb n}$. We have also introduced the notations
\be
\Delta_{\text{sd}}=\vert J_{\text{sd}}\vert S, \qquad \varsigma=\sign{J_{\text{sd}}},
\e
where $\Delta_{\text{sd}}$ has a meaning of half of the exchange interaction-induced splitting (in the absence of SOC).

\begin{figure}[t!]
\includegraphics[width=0.65\columnwidth]{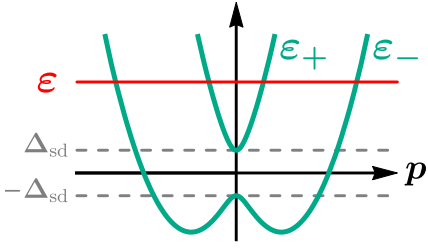}
\caption{Guide for an eye: spectrum of the homogeneous system of conduction electrons with a fixed direction of magnetization. Note that the actual spectrum is not isotropic, and the two subbands may even touch each other. We restrict the analysis to the case of $\varepsilon>\Delta_{\text{sd}}$. For the latter, both subbands are always partly filled.}
\label{fig::spectrum}
\end{figure}

If the chemical potential $\varepsilon$ exceeds the value of $\Delta_{\text{sd}}$, both subbands are always partly filled~\cite{KimMATH2016}. Below, we focus solely on the latter case, which is schematically illustrated in Fig.~\ref{fig::spectrum}. Note that the spectrum is not isotropic. Moreover, for finite values of $\sin{\theta}$, separation of the two subbands diminishes and they may even touch each other.

In what follows, we also find it convenient to introduce the energy scale $\Delta_{\text{so}}=\vert\alpha_{\text{\tiny R}}\vert\sqrt{2m\ep}$, which is equal to half of the spin-orbit coupling-induced splitting of the branches (for vanishing $\Delta_{\text{sd}}$).

\subsection{Roots of dispersion relation}
\label{sec::roots}
Now let us analyze the roots of the dispersion of Eq.~(\ref{spectrum}). Using, for example, Ref.~[\onlinecite{1922Rees}], one can show that, under the assumption $\varepsilon>\Delta_{\text{sd}}$, the quartic function $(\varepsilon_+(\bb p)-\varepsilon)(\varepsilon_-(\bb p)-\varepsilon)$ of the absolute value of momentum $p$ always has four real roots: two positive and two negative. The former two define the angle-dependent Fermi momenta $p_{\pm}$ corresponding to $\varepsilon_{\pm}$ branches. The four roots are distinct in all cases, except one. Namely, when $\bb n_\perp=0$ (i.e.,~when $\sin{\theta}=1$) and $\Delta_{\text{so}}=\Delta_{\text{sd}}$, the subbands touch each other. We will not consider this particular case.

Using the notation $p_{\pm,\text{neg}}$ for the negative roots, we have
\be
p_- > p_+ > 0 > p_{+,\text{neg}}>p_{-,\text{neg}},
\e
where
\begin{subequations}
\label{p_and_pneg}
\begin{gather}
\label{p}
p_\mp=\frac{1}{2}\left(\sqrt{2u}\pm\sqrt{-2u-2q-r\sqrt{2/u}}\right),
\\
\label{pneg}
p_{\pm,\text{neg}}=\frac{1}{2}\left(-\sqrt{2u}\pm\sqrt{-2u-2q+r\sqrt{2/u}}\right)\, ,
\end{gather}
\end{subequations}
$u>0$ is the largest root of the resolvent cubic
\be
\label{resolvent}
u^3+q u^2-(s-q^2/4)u-r^2/8,
\e
while the parameters $q$, $s$, and $r$ are given by
\begin{subequations}
\label{quartic_qsr}
\begin{gather}
\label{quartic_qs}
q=-4m(\varepsilon+m\alpha_{\text{\tiny R}}^2), \qquad s=(2m)^2(\varepsilon^2-\Delta_{\text{sd}}^2),
\\
\label{quartic_r}
r=8m^2\alpha_{\text{\tiny R}}\varsigma\Delta_{\text{sd}}\sin{\theta}\sin{\varphi}.
\end{gather}
\end{subequations}

It is straightforward to see, from Eqs.~(\ref{quartic_qsr}), that the dependence on the momentum angle enters Eq.~(\ref{resolvent}) only via the parameter $r^2$. As a result, the quantity $u$ may only depend on $\sin^2{\varphi}$ and other parameters of the model that are $\varphi$ independent. This will play an important role below.

For $\alpha_{\text{\tiny R}}=0$ (vanishing SOC), $\Delta_{\text{sd}}=0$ (nonmagnetic limit), or $\bb n=\bb n_{\perp}$ (perpendicular-to-the-plane magnetization) situation with the roots becomes less complex. In these cases, $(\varepsilon_+(\bb p)-\varepsilon)(\varepsilon_-(\bb p)-\varepsilon)$ is biquadratic (with respect to $p$) and $p_\pm=-p_{\pm,\text{neg}}$, as one can also see directly from Eqs.~(\ref{p_and_pneg}). Furthermore, the Fermi momenta $p_{\pm}$, then, are angle independent, while their values yield the relations
\begin{subequations}
\label{zero_r_poles}
\begin{align}
\label{zero_SOC_poles}
p_{\pm}^2=2m\left[\varepsilon\mp\Delta_{\text{sd}}\right],\,\quad &\text{for $\alpha_{\text{\tiny R}}=0$},
\\
\label{zero_J_poles}
p_{\pm}^2=2m\left[\varepsilon+m\alpha_{\text{\tiny R}}^2
\mp \lambda(0)\right],\,\quad &\text{for $\Delta_{\text{sd}}=0$},
\\
\label{nperp_poles}
p_{\pm}^2=2m\left[\varepsilon+m\alpha_{\text{\tiny R}}^2
\mp \lambda(\Delta_{\text{sd}})\right],\,\quad &\text{for $\bb n=\bb n_{\perp}$},
\end{align}
\end{subequations}
where $\lambda(\Upsilon)=\sqrt{\Upsilon^2+2\varepsilon m\alpha_{\text{\tiny R}}^2+m^2\alpha_{\text{\tiny R}}^4}$.

\section{Disorder averaging}
\label{sec::disorder}
Having analysed the spectrum of the ``clean'' homogeneous system, we can proceed with the inclusion of the disorder. In what follows, we assume $\varepsilon_0 \tau\gg 1$, where $\varepsilon_0$ is the difference between the Fermi energy $\varepsilon$ and the closest band edge. We start with a calculation of the self-energy in the first Born approximation.
\subsection{Self-energy}
\label{sec::self-energy}
According to Eq.~(\ref{V_correlator}), the self-energy is defined as
\be
\label{Born_general}
\Sigma^{R,A}(\bb r)=(\hbar^2/m\tau)\,\mathcal G^{R,A}(\bb r, \bb r),
\e
with the Green's function $\mathcal G^{R,A}$ of Eq.~(\ref{Gmathcal}). It should be explicitly pronounced that $\Sigma^{R,A}(\bb r)$ may have a spatial dependence originating in the spatial dependence of~$\bb n(\bb r)$. However, as we are about to see, the first spatial gradients of magnetization do not affect the self-energy in the model under consideration.

Disregarding the ``real'' part of the self-energy that should be included in the renormalized value of the chemical potential, we focus only on the calculation of $\im{\Sigma\,(\bb r)}=-i[\Sigma^R(\bb r)-\Sigma^A(\bb r)]/2$. By substituting the expansion of Eq.~(\ref{G_gradient}) into Eq.~(\ref{Born_general}), switching to momentum representation, and symmetrizing the result we obtain
\be
\label{Sigma_general}
\im{\Sigma\,(\bb r)}=\Sigma^{(0)}+
\sum_{\gamma\delta}{\left\{(\bb r-\bb r_\ast)_\gamma\,\Sigma^{(1)}_{\delta}+\Sigma^{(2)}_{\gamma\delta}\right\}\,\nabla_\gamma n_\delta(\bb r_\ast)},
\e
with
\begin{subequations}
\label{Sigma}
\begin{gather}
\label{Sigma_0}
\Sigma^{(0)}=\frac{1}{2 i m\tau}
\int{\frac{d^2 p}{(2\pi)^2}
\left(G^R-G^A\right)},
\\
\label{Sigma_1}
\Sigma^{(1)}_{\delta}=\frac{\varsigma\Delta_{\text{sd}}}{2i m\tau}
\int
\frac{d^2 p}{(2\pi)^2}
\Bigl(
G^{R}\sigma_\delta\,G^{R}-G^{A}\sigma_\delta\,G^{A}
\Bigr),
\\
\begin{multlined}
\label{Sigma_2}
\Sigma^{(2)}_{\gamma\delta}=\frac{\varsigma\Delta_{\text{sd}}\hbar}{4 m\tau}
\int
\frac{d^2 p}{(2\pi)^2}
\Bigl(
G^{R}\sigma_\delta\,G^{R}\,v_\gamma\,G^{R}-
\\
G^{R}\,v_\gamma\,G^{R}\sigma_\delta\,G^{R}+\text{h.c.}
\Bigr),
\end{multlined}
\end{gather}
\end{subequations}
where ``h.c.'' denotes Hermitian conjugate, $G^{R,A}$ is the Green's function of Eq.~(\ref{G}) in momentum representation,
\be
\label{G_def}
G^{R,A}\hspace{-0.1ex}=\hspace{-0.1ex}\frac{\varepsilon-p^2/2m \hspace{-0.1ex}+\hspace{-0.1ex} \alpha_{\text{\tiny R}}\, [\bb{p}\times\bb{\sigma}]_z  \hspace{-0.1ex}+\hspace{-0.1ex} \varsigma\Delta_{\text{sd}}\,\bb n(\bb r_\ast)\hspace{-0.3ex}\cdot\hspace{-0.3ex}\bb{\sigma}}{(\varepsilon-\varepsilon_+(\bb p)\pm i0)(\varepsilon-\varepsilon_-(\bb p)\pm i0)},
\e
and $\bb v=\partial H/\partial \bb p$ is the velocity operator. In Eqs.~(\ref{Sigma}), $\Sigma^{(0)}$ defines the scattering time (for uniform magnetization), $\Sigma^{(1)}$ corresponds to the renormalization of the gradient term on the right hand side of Eq.~(\ref{H_gradient}), while $\Sigma^{(2)}$ determines the possible dependence of the scattering time on the first spatial gradients of magnetization. 

To proceed, we take advantage of the additional symmetrization of the integrands with respect to the transformation~\cite{angle_integration} $\varphi\to\pi-\varphi$ and observe that, in the first Born approximation, integration over the absolute value of momentum, in Eqs.~(\ref{Sigma}), is reduced to a calculation of residues at $p=p_\pm$. Using Eqs.~(\ref{p_and_pneg}), we, then, get
\begin{multline}
\Sigma^{(0)}=-\frac{1}{2\tau}\int_0^{2\pi}\frac{d\varphi}{2\pi}\bigl[1+r W_1+r W_2\,\bb n(\bb r_\ast)\cdot\bb \sigma
\\
+W_3\,\bb n_\parallel(\bb r_\ast)\cdot\bb \sigma\sin{\varphi}\bigr],
\end{multline}
where $W_i=W_i\left(r^2,u\,(r^2)\right)$ are some functions of the parameter $r^2$ and $\varphi$-independent parameters of the model. Since $r\propto\sin{\varphi}$ and, obviously, all integrals of the form $\int_0^{2\pi}{W\,(\sin^2{\varphi})\sin{\varphi}\,d\varphi}$ vanish for arbitrary function~$W$, we obtain a particularly simple result for the constant part of the self-energy,
\be
\label{Sigma_calculated}
\Sigma^{(0)}=-1/2\tau.
\e

Similar, but more lengthy, analysis shows that each component of $\Sigma^{(1)}$ and $\Sigma^{(2)}$ is equal to zero. Therefore, there exists no renormalization of the gradient term of the Hamiltonian $\mathcal H$ as well as no scattering time dependence on the first magnetization gradients. The self-energy, in the first Born approximation, is found as
\be
\label{self_energy}
\Sigma^{R,A}(\bb r)=\mp i/2\tau.
\e

\subsection{Kubo formula for STT}
\label{sec::STT_Kubo}
As was outlined in Sec.~\ref{sec::SPproblem}, the generalized torque $\bb T(\bb r_0)$ of Eq.~(\ref{torque_gen}), at a certain position $\bb r_0$ in space, is defined as a disorder-averaged mean value of the operator $(J_{\text{sd}}A/\hbar)\delta_{\bb r_0}\hat{\bb T}$, where $\delta_{\bb r_0}= \delta(\bb r-\bb r_0)$. At zero temperature, the linear response~\cite{comment_on_Kubo} of $T_\alpha(\bb r_0)$ to the zero frequency electric field $\bb E$ is given by the standard Kubo expression
\be
\label{E_Kubo}
\frac{e\hbar}{2\pi}\frac{J_{\text{sd}}A}{\hbar}\left\langle
\tr{\left[\mathcal G^A\delta_{\bb r_0}\hat{T}_\alpha\,\mathcal G^R{\bb v}\right]}\bb E
\right\rangle,
\e
where $\bb v=\partial \mathcal H/\partial\bb p$ is the velocity operator, $\tr{}$ stands for the operator trace, and angular brackets represent the disorder averaging.

From Eq.~(\ref{E_Kubo}), we can further deduce the Kubo formula for spin-transfer torques. In order to do that, we substitute the expansion of Eq.~(\ref{G_gradient}) into Eq.~(\ref{E_Kubo}) and collect all terms proportional to $\nabla_\gamma n_\delta(\bb r_\ast)$. Then we switch to momentum representation and perform spatial averaging of torque on the scale of transport mean free path in the vicinity of $\bb r=\bb r_0$. In the noncrossing approximation, this leads to the general formula for the STT tensor,
\begin{multline}
\label{STT_tensor_def}
\mathcal{T}^{\text{STT}}_{\alpha\beta\gamma\delta}=
\frac{e\Delta_{\text{sd}}^2 A}{2\pi\hbar S}
\int
\frac{d^2 p}{(2\pi)^2}
\\
\times i\mtr{\Bigl[g^A\,\sigma_{\delta}\,g^A\,v_{\gamma}\,g^A\,\hat{T}^{\text{vc}}_\alpha\,g^R\,v^{\text{vc}}_{\beta}
-\text{h.c.}\Bigr]},
\end{multline}
where the superscript ``$\text{vc}$'' marks the vertices corrected with the impurity ladders, the notation $\mtr$ refers to the matrix trace, and
\be
\label{g}
g^{R,A}=\langle G^{R,A}\rangle=(\varepsilon-H\pm i/2\tau)^{-1}
\e
is the disorder-averaged Green's function of the homogeneous system. In Eq.~(\ref{g}), we have used the result for the self-energy obtained in Sec.~\ref{sec::self-energy}.

\begin{figure}[t]
\includegraphics[width=1\columnwidth]{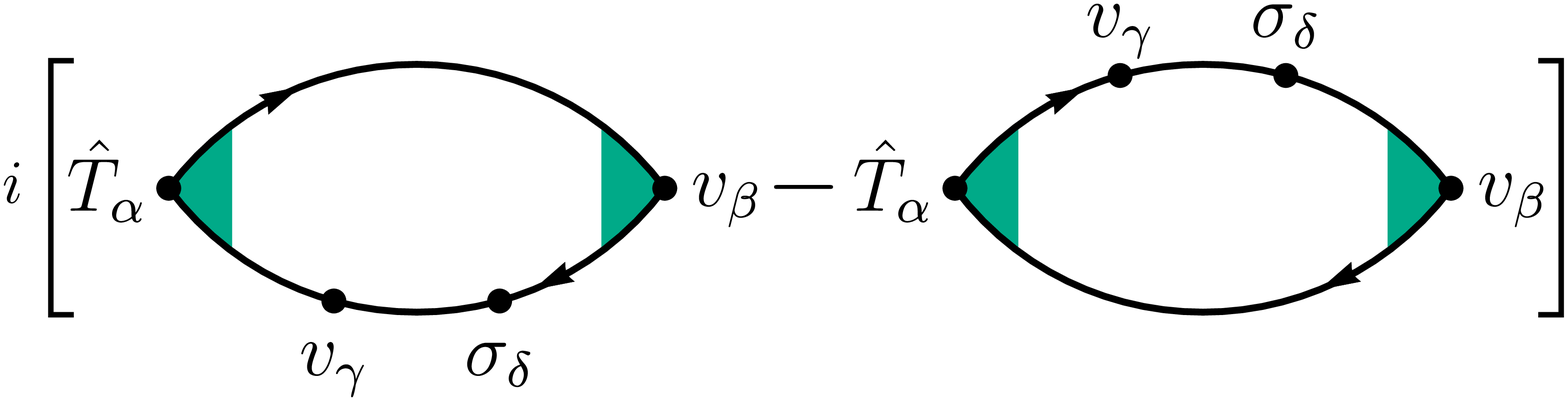}
\caption{Diagrammatic representation of the STT tensor $\mathcal{T}^{\text{STT}}_{\alpha\beta\gamma\delta}$ of Eq.~(\ref{STT_tensor_def}). Solid lines correspond to the disorder-averaged Green's functions $g^{R,A}$. Vertex corrections (impurity ladders) are represented by green fillings.}
\label{fig::STT-tensor}
\end{figure}

The expression of Eq.~(\ref{STT_tensor_def}) is represented diagrammatically in Fig.~\ref{fig::STT-tensor}. We note that similar diagrams have been used in Ref.~[\onlinecite{Kohno2006}] to compute STT in a 3D FM, in the absence of SOC, and in Ref.~[\onlinecite{Sakai2014}] to study STT for the model of massive Dirac fermions.


\subsection{Kubo formula for GD and ESR}
Similarly, from the zero frequency linear response~\cite{comment_on_Kubo} of $T_\alpha(\bb r_0)$ to the time derivative~of~$\bb n$,
\be
\label{n_Kubo}
\frac{J_{\text{sd}}S\hbar}{2\pi}\frac{J_{\text{sd}}A}{\hbar}\left\langle
\tr{\left[\mathcal G^A\delta_{\bb r_0}\hat{T}_\alpha\,\mathcal G^R \bb \sigma\right]}\partial_{t}\bb n
\right\rangle,
\e
one may derive the formula for the GD tensor of Eq.~(\ref{GD_def}),
\be
\label{GD_tensor_def}
\mathcal T^{\text{GD}}_{\alpha\delta}=
\frac{\Delta_{\text{sd}}^2 A}{2\pi\hbar^2 S}
\int{
\frac{d^2 p}{(2\pi)^2}
\mtr{\Bigl[g^A \,\hat{T}^{\text{vc}}_\alpha\, g^R\,\sigma_\delta\Bigr]}},
\e
where, according to the definition of $\bb T^{\text{GD}}$, spatial dependence of $\bb n$ is completely disregarded.

Note that $\bb n$, $\nabla_\gamma n_\delta$, and $\partial_{t}\bb n$ in Eqs.~(\ref{STT_and_GD_def}),~(\ref{STT_tensor_def}), and (\ref{GD_tensor_def}) are all taken at $\bb r=\bb r_0$. From now on, we consistently omit the argument of all these functions.

\subsection{Relation between $\bb T^{\text{GD}}$ and vertex corrections to the torque operator $\hat{\bb T}$}
\label{sec::GD_to_T}
\begin{figure*}[ht!]
\includegraphics[width=0.97\textwidth]{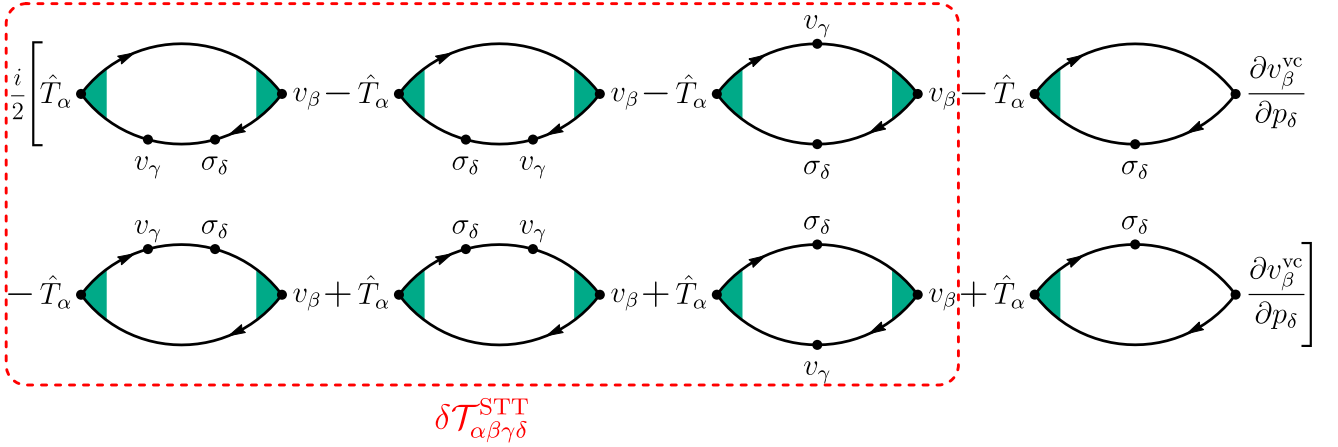}
\caption{Another diagrammatic representation of the STT tensor $\mathcal{T}^{\text{STT}}_{\alpha\beta\gamma\delta}$, given by Eq.~(\ref{STT_tensor_sym_1}). 
Six diagrams encircled by the dashed line define the $\delta\mathcal{T}^{\text{STT}}_{\alpha\beta\gamma\delta}$ tensor of Eq.~(\ref{delta_STT_tensor}) that vanishes for any direction of $\bb n$ provided $\varepsilon>\Delta_{\text{sd}}$. Solid lines correspond to the disorder-averaged Green's functions $g^{R,A}$. Vertex corrections (impurity ladders) are represented by green fillings.}
\label{fig::STT-tensor_long}
\end{figure*}
Vertex corrected torque operator that enters both Eqs.~(\ref{STT_tensor_def}) and (\ref{GD_tensor_def}) can be expressed with the help of vertex corrected Pauli matrices. One can infer the latter from the ``matrix of one dressing'' $\mathcal M$, whose elements
\be
\label{M_def}
\mathcal M_{i j}=\frac{1}{2m\tau}\int{
\frac{d^2 p}{(2\pi)^2}
\mtr{\Bigl[g^A \,\sigma_i\, g^R\,\sigma_j\Bigr]}}
\e
are the coordinates (in the basis $\{\sigma_x, \sigma_y, \sigma_z\}$) of the operator $\sigma_i$ dressed with a single impurity line. We note that, in the model considered, vertex corrected Pauli matrices $\sigma_i^{\text{vc}}$ appear to have zero trace if $\varepsilon>\Delta_{\text{sd}}$. This is a direct consequence of the fact that the self-energy in Eq.~(\ref{self_energy}) is scalar. Hence, $\{\sigma_x, \sigma_y, \sigma_z\}$ is, indeed, a proper basis for the operators $\sigma_i^{\text{vc}}$.

Matrix representation of the operator \mbox{$\hat{\bb T}=\bb \sigma\times\bb n$}, with respect to this basis, is defined as
\be
\label{U_def}
\hat{T}_i=\sum\limits_j{U_{i j}\,\sigma_j}, \qquad
U=
\begin{pmatrix}
0& n_z & -n_y \\
-n_z & 0 &n_x \\
n_y & -n_x & 0
\end{pmatrix}.
\e
Since, obviously,
\be
\hat{T}^{\text{vc}}_i=\sum\limits_j{U_{i j}\,\sigma_j^{\text{vc}}},
\e
we can see, from Eq.~(\ref{M_def}), that the geometric series
\be
\label{T_matrix_def}
\mathcal T=U(\mathcal M+\mathcal M^2+\cdots)=U\mathcal M(I-\mathcal M)^{-1},
\e
provides the matrix representation of vertex corrections to the torque operator. Moreover, from Eq.~(\ref{GD_tensor_def}), it is evident that the GD tensor is, in fact, determined by the same matrix $\mathcal T$,
\be
\label{GD_to_T}
\mathcal T^{\text{GD}}_{\alpha\delta}=
\frac{\Delta_{\text{sd}}^2 A m\tau}{\pi\hbar^2 S}\mathcal T_{\alpha\delta}.
\e

\subsection{Crossing diagrams}
It has been demonstrated recently that the diagrams with two crossing impurity lines may contribute to such quantities as the anomalous Hall effect\cite{2015AdoEPL, 2016AdoPRL, Ado2017PRBcrossing}, the spin Hall effect\cite{2016MilletariSHE}, and the Kerr effect\cite{Konig2017} in the same leading order with respect to the small parameter $(\varepsilon_0\tau)^{-1}$, as the conventional noncrossing approximation does. Scattering mechanisms associated with these diagrams, in general, should affect spin torques and damping as well.

In the present study we, however, completely disregard the crossing diagrams, as being significantly more difficult to calculate. At the same time, preliminary analysis shows that the related additional contributions to STT, GD, and ESR are parametrically different from the present results and that, for $\varepsilon\gg\Delta_{\text{sd}}$, they are negligible.


\section{Relation between STT, GD, and ESR}
\label{sec::relation}
\subsection{Symmetrization of STT diagrams}
Calculation of spin-transfer torques can be performed with the help of Eq.~(\ref{STT_tensor_def}) directly. Such brute-force calculation has been originally performed by us. We have, however, subsequently found a shortcut that makes it possible not only to obtain the same results in a much more concise manner but also to establish a general relation between $\mathcal{T}^{\text{STT}}$ and $\mathcal{T}^{\text{GD}}$ tensors. This alternative approach takes a reformulation of the result of Eq.~(\ref{STT_tensor_def}) in a more symmetric form.

We apply the identity $g^A\,v_{\gamma}\,g^A=\partial g^A/\partial p_\gamma$ in Eq.~(\ref{STT_tensor_def}) and perform integration by parts. Then, we take a half-sum of the result obtained and the original expression of Eq.~(\ref{STT_tensor_def}). This leads to the formula
\begin{multline}
\label{STT_tensor_sym_1}
\mathcal{T}^{\text{STT}}_{\alpha\beta\gamma\delta}=
\delta\mathcal T^{\text{STT}}_{\alpha\beta\gamma\delta}+\frac{e\Delta_{\text{sd}}^2 A}{2\pi\hbar S}
\int
\frac{d^2 p}{(2\pi)^2}
\\
\times \frac{i}{2}\mtr{\biggl[-g^A\,\sigma_{\delta}\,g^A\,\hat{T}^{\text{vc}}_\alpha\,g^R\,\frac{\partial v^{\text{vc}}_{\beta}}{\partial p_\gamma}
-\text{h.c.}\biggr]},
\end{multline}
where the first term on the right-hand side
\begin{multline}
\label{delta_STT_tensor}
\delta\mathcal{T}^{\text{STT}}_{\alpha\beta\gamma\delta}=
\frac{e\Delta_{\text{sd}}^2 A}{2\pi\hbar S}
\int
\frac{d^2 p}{(2\pi)^2}
\\
\frac{i}{2}\mtr\Bigl[g^A\,\sigma_{\delta}\,g^A\,v_{\gamma}\,g^A\,\hat{T}^{\text{vc}}_\alpha\,g^R\,v^{\text{vc}}_{\beta} -
g^A\,v_{\gamma}\,g^A\,\sigma_{\delta}\,g^A\,\hat{T}^{\text{vc}}_\alpha\,g^R\,v^{\text{vc}}_{\beta}
\\
-g^A\,\sigma_{\delta}\,g^A\,\hat{T}^{\text{vc}}_\alpha\,g^R\,v_{\gamma}\,g^R\,v^{\text{vc}}_{\beta}-\text{h.c.}\Bigr].
\end{multline}
is illustrated schematically in Fig.~\ref{fig::STT-tensor_long} by a group of encircled diagrams. The remaining two diagrams in Fig.~\ref{fig::STT-tensor_long} correspond to the second term on the right-hand side of Eq.~(\ref{STT_tensor_sym_1}). We will see below that, in fact, the entire tensor $\delta\mathcal{T}^{\text{STT}}$ does vanish.

\subsection{Relation between $\bb T^{\text{STT}}$ and vertex corrections to the torque operator $\hat{\bb T}$}
As was argued in Ref.~\onlinecite{AdoSOT2017} on the basis of perturbative expansions, the velocity operator $\bb v=\bb p/m-\alpha_{\text{\tiny R}}[\bb e_z\times\bb\sigma]$, corrected by an impurity ladder, has a particularly simple form in the present model,
\be
\label{VCtoV}
\bb v^{\text{vc}}=\bb p/m.
\e
A formal proof of this statement that does not refer to any perturbative expansion is presented in Appendix~\ref{sec::VCtoV}. Interestingly, Eq.~(\ref{VCtoV}) also allows to make a spin-orbit torque (SOT) calculation extremely concise. We provide a brief discussion of this matter in the same Appendix~\ref{sec::VCtoV}.

It is important that the momentum operator $\bb p$, as well as $\bb v^{\text{vc}}$, commutes with the Green's function $g^{R,A}$. In Appendix~\ref{sec::deltaT_vanishes}, we demonstrate that this is sufficient for the entire tensor $\delta\mathcal{T}^{\text{STT}}$ to vanish. As a result, $\mathcal{T}^{\text{STT}}$ is determined by the second term on the right hand side of Eq.~(\ref{STT_tensor_sym_1}) alone. Computation of the this term is facilitated by the relation
\be
\label{derivative_of_vVC}
\partial v^{\text{vc}}_{\beta}/\partial p^{\phantom{\text{vc}}}_\gamma=\delta_{\beta\gamma}/m,
\e
where $\delta_{q_1 q_2}$ is Kronecker delta. With the help of the above, the STT tensor of Eq.~(\ref{STT_tensor_sym_1}) readily simplifies to
\begin{multline}
\label{STT_tensor_sym_2}
\mathcal{T}^{\text{STT}}_{\alpha\beta\gamma\delta}=
\delta_{\beta\gamma}\frac{e\Delta_{\text{sd}}^2 A}{2\pi\hbar S m}
\int
\frac{d^2 p}{(2\pi)^2}
\\
\times \frac{i}{2}\mtr{\Bigl[-g^A\,\sigma_{\delta}\,g^A\,\hat{T}^{\text{vc}}_\alpha\,g^R
-\text{h.c.}\Bigr]},
\end{multline}
since, as we have mentioned, $\delta\mathcal{T}^{\text{STT}}=0$.

Employing the Hilbert's identity for the Green's functions of Eq.~(\ref{g}),
\be
\label{Hilbert}
g^A-g^R=g^R\,(i/\tau)\,g^A,
\e
we can further reduce~\cite{comment_on_Kubo} Eq.~(\ref{STT_tensor_sym_2}) to the formula
\be
\label{STT_tensor_sym_3}
\mathcal{T}^{\text{STT}}_{\alpha\beta\gamma\delta}=
\delta_{\beta\gamma}\frac{e\Delta_{\text{sd}}^2 A\tau}{2\pi\hbar S m}
\int{
\frac{d^2 p}{(2\pi)^2}\mtr{\Bigl[
g^A\,\hat{T}^{\text{vc}}_\alpha\,g^R\,\sigma_{\delta}\Bigr]}},
\e
which resembles very closely the formula of Eq.~(\ref{GD_tensor_def}) for the GD tensor. The result of Eq.~(\ref{STT_tensor_sym_3}) can also be expressed in terms of the matrix $\mathcal T$ as
\be
\label{STT_to_T}
\mathcal{T}^{\text{STT}}_{\alpha\beta\gamma\delta}=\delta_{\beta\gamma}
\frac{e\Delta_{\text{sd}}^2 A\tau^2}{\pi\hbar S}\mathcal T_{\alpha\delta},
\e
where we have again used the argumentation of Sec.~\ref{sec::GD_to_T}.

\subsection{Relation between $\bb T^{\text{STT}}$ and $\bb T^{\text{GD}}$}
It can now be seen that both $\bb T^{\text{STT}}$ and $\bb T^{\text{GD}}$ vectors turn out to be fully defined by the matrix of vertex corrections $\mathcal T$ to the torque operator. Moreover, comparison of Eq.~(\ref{GD_to_T}) and Eq.~(\ref{STT_to_T}) reveals a remarkable direct connection between the STT and GD tensors,
\be
\mathcal{T}^{\text{STT}}_{\alpha\beta\gamma\delta}=\delta_{\beta\gamma}
\frac{e \hbar\tau}{m}\mathcal T^{\text{GD}}_{\alpha\delta},
\e
which is one of the central results of the paper.

According to the definitions of Eqs.~(\ref{STT_and_GD_def}), the established relation between the two tensors indicates that all quantities of interest (STT, GD, and ESR) may be related to the action of a single linear \mbox{operator $\Xi$,}
\be
\label{Xi}
\bb T^{\text{STT}}=\Xi\left[\partial_{\bb v}\bb n\right], \qquad \bb T^{\text{GD}}=\Xi\left[\partial_{t}\bb n\right],
\e
on one of the vectors, $\partial_{\bb v}\bb n$ or $\partial_t\bb n$. We remind here the short-handed notations for the directional spatial derivative~\cite{comment_on_derivative} $\partial_{\bb v}=(\bb v_{\text{d}}\cdot\bb\nabla)$ and for the classical drift velocity of conduction electrons $\bb v_{\text{d}}=e\bb E\hbar\tau/m$.

The matrix of the operator $\Xi$ coincides with the matrix $\mathcal T^{\text{GD}}$, being also proportional to the matrix~$\mathcal T$ [see Eqs.~(\ref{GD_def}),~and~(\ref{GD_to_T})]. In the next section we obtain the general form of the latter and then use it to derive the exact vector forms of $\bb T^{\text{STT}}$ and $\bb T^{\text{GD}}$.

\section{Vector forms}
\label{sec::vector_forms}
\subsection{Matrix gauge transformation}
In order to establish the structure of the operator~$\Xi$, it should be first noted that the constraint $\bb n^2 \equiv 1$ is responsible for an essential freedom in the definition of~$\mathcal T$. For an arbitrary operator of differentiation $\partial$, we have
\be
\label{wisdom}
\frac{1}{2}\partial \bb n^2=\sum_{\delta}{n_{\delta}\, \partial n_\delta}=0.
\e
Therefore, the left hand sides of
\begin{subequations}
\label{STTandGD_def_new}
\begin{align}
\label{STT_def_new}
T^{\text{STT}}_\alpha &=\frac{\Delta_{\text{sd}}^2 A m\tau}{\pi\hbar^2 S}
\sum_{\delta}{\mathcal T_{\alpha\delta}\, \partial_{\bb v} n_\delta},
\\
\label{GD_def_new}
T^{\text{GD}}_\alpha &=\frac{\Delta_{\text{sd}}^2 A m\tau}{\pi\hbar^2 S}
\sum_{\delta}{\mathcal T_{\alpha\delta}\, \partial_{\,t \phantom{\bb v}\nphantom{\,t}} n_\delta},
\end{align}
\end{subequations}
remain invariant under the addition of the matrix row $\mathcal R=(n_x, n_y, n_z)$, with an arbitrary coefficient, to any of the rows of the matrix $\mathcal T$. In other words, the transformation $\mathcal T\to \mathcal T_X$ does not change $\bb T^{\text{STT}}$ and $\bb T^{\text{GD}}$, provided
\be
\label{transformation_def}
\mathcal T_{X}=\mathcal T+ X \mathcal R,
\e
with any matrix column $X=(X_1, X_2, X_3)^T$.


\subsection{Vector structure of $\bb T^{\text{STT}}$ and $\bb T^{\text{GD}}$}
The matrix $\mathcal T$ is defined in Eq.~(\ref{T_matrix_def}) with the help of the matrix $\mathcal M$. The latter is determined by the disorder-averaged Green's function which, in momentum representation, takes the form
\be
\label{g_in_momentum_space}
g^{R,A}=\frac{\varepsilon\pm i/2\tau-p^2/2m + \alpha_{\text{\tiny R}}\, [\bb{p}\times\bb{\sigma}]_z  + \varsigma\Delta_{\text{sd}}\,\bb n\cdot\bb{\sigma}}{(\varepsilon-\varepsilon_+(\bb p)\pm i/2\tau)(\varepsilon-\varepsilon_-(\bb p)\pm i/2\tau)}.
\e
Using Eq.~(\ref{g_in_momentum_space}), one can prove that $\mathcal M$, in general, is expressed as a linear combination of six matrices,
\be
I,\,\,\, P,\,\,\, U,\,\,\, U^2,\,\,\, P\,UP,\,\,\, P\,U^2P,
\e
where $U$ is introduced in Eq.~(\ref{U_def}) and $P=\diag{(1,1,0)}$ is a diagonal matrix. In Appendix~\ref{sec::structure_of_M}, we demonstrate how the components of this decomposition can be calculated for $\bb n\neq\bb n_{\perp}$.

Then, in Appendix~\ref{sec::structure_of_M^k}, we show that any power of $\mathcal M$ retains the same structure. It immediately follows that the matrix $\mathcal T=U(\mathcal M+\mathcal M^2+\cdots)$ can be represented as 
\be
\label{Tsymm}
\mathcal T =c_1 U+c_2 U P+c_3 U^2+c_4 U^3+c_5 U P\, U P+c_6 U P\, U^2 P,
\e
where $c_i$ are some dimensionless scalar functions.

The representation of Eq.~(\ref{Tsymm}) can be substantially simplified with the use of the matrix gauge transformation described in the previous section. Namely, by taking advantage of the directly verifiable relations
\begin{subequations}
\begin{gather}
\label{gauge_identity_1}
U^2=\mathcal R^T \mathcal R-I,\qquad U^3=-U,\\
U P\,U P=(I-P)\mathcal R^T \mathcal R-n_z^2 I,\\
U P\, U^2 P=UP\,\mathcal R^T \mathcal R-U P+n_z^2 U(I-P)
\end{gather}
\end{subequations}
we find that the choice of the gauge
\be
\label{definiton_of_X_tilded}
{\widetilde X}=-\left[c_3 I + c_5 (I-P) + c_6 U P\right]\mathcal R^T,
\e
for the transformation $\mathcal T\to \mathcal T_{\widetilde X}\equiv\widetilde{\mathcal T}$, leads to
\be
\label{definiton_of_T_tilded}
\widetilde{\mathcal T}=t_0\,I+t_\parallel\, U P+t_\perp\, U(I-P),
\e
or, more explicitly, to
\be
\widetilde{\mathcal T}=
\begin{pmatrix}
t_0 & n_z t_{\parallel} & -n_y t_{\perp}\\
-n_z t_{\parallel} & t_0 & n_x t_{\perp}\\
n_y t_{\parallel} & -n_x t_{\parallel} & t_0
\end{pmatrix},
\e
where the quantities $t_i$ are related to the matrix $\mathcal T$ by means of the relations
\begin{subequations}
\label{t_i_def}
\begin{gather}
\label{t_0_def}
t_0=-c_3-c_5 n_z^2,\\
\label{t_par_def}
t_{\parallel}=c_1+c_2-(c_4+c_6),\\
\label{t_perp_def}
t_{\perp}=c_1-c_4+c_6 n_z^2.
\end{gather}
\end{subequations}

Replacing $\mathcal T$ with $\widetilde{\mathcal T}$ in Eqs.~(\ref{STTandGD_def_new}),
\begin{subequations}
\begin{align}
\label{STT_def_final}
T^{\text{STT}}_\alpha &=\frac{\Delta_{\text{sd}}^2 A m\tau}{\pi\hbar^2 S}
\sum_{\delta}{\widetilde{\mathcal T}_{\alpha\delta}\, \partial_{\bb v} n_\delta},
\\
\label{GD_def_final}
T^{\text{GD}}_\alpha &=\frac{\Delta_{\text{sd}}^2 A m\tau}{\pi\hbar^2 S}
\sum_{\delta}{\widetilde{\mathcal T}_{\alpha\delta}\, \partial_{\,t \phantom{\bb v}\nphantom{\,t}} n_\delta},
\end{align}
\end{subequations}
we observe that the operator $\Xi$ in Eq.~(\ref{Xi}) is represented by three dimensionless quantities $\xi_0$, $\xi_\parallel$, $\xi_\perp$, such that
\be
\label{xi_from_t}
\xi_i=\frac{\Delta_{\text{sd}}^2 A m\tau}{\pi\hbar^2 S}t_i,
\e
while the vector structure of $\bb T^{\text{STT}}$ and $\bb T^{\text{GD}}$ is, indeed, provided by the formulas
\begin{align*}
\bb T^{\text{STT}}&=
\xi_0 \partial_{\bb v}\bb n
-\xi_{\parallel}[\bb n\times \partial_{\bb v}\bb{n}_{\parallel}]-\xi_{\perp}[\bb n \times \partial_{\bb v}\bb{n}_{\perp}],
\\
\bb T^{\text{GD}}&=
\xi_0 \partial_{\,t \phantom{\bb v}\nphantom{\,t}}\bb n
-\xi_{\parallel}[\bb n\times \partial_{\,t \phantom{\bb v}\nphantom{\,t}}\bb n_{\parallel}]
-\xi_{\perp}[\bb n \times \partial_{\,t \phantom{\bb v}\nphantom{\,t}}\bb n_{\perp}],
\end{align*}
announced in the introductory part. With some remarks, they remain valid for $\bb n=\bb n_{\perp}$ as well. We consider this specific case separately, in Sec.~\ref{sec::out_of_plane}. 

In the next section, we derive closed-form results for $\xi_0$, $\xi_\parallel$, and $\xi_\perp$, in two particular regimes. Afterwards, we find asymptotic expansions of these functions in either small $\alpha_{\text{\tiny R}}$ or in small $\Delta_{\text{sd}}$. All the obtained results are collected in Table~\ref{table} and represented in Fig.~\ref{fig::plots} alongside with the corresponding numerical curves.

\section{Closed-forms}
\label{sec::closed_forms}
The analysis of $\mathcal T^{\text{STT}}$ and $\mathcal T^{\text{GD}}$ tensors, as has been pointed out, reduces to integration in Eq.~(\ref{M_def}) and subsequent matrix arithmetics. Unfortunately, for arbitrary direction of magnetization, the results cannot be expressed in terms of elementary functions. For example, for $\bb n_\perp=0$, Eq.~(\ref{M_def}) already involves elliptic integrals. The complexity is caused, primarily, by the angle dependence of the dispersion relation roots $p_\pm$, $p_{\pm,\text{neg}}$ of Eqs.~(\ref{p_and_pneg}). Additional complications arise due to the fact that all four roots are distinct. 

On the other hand, if the parameter $r$ defined in Eq.~(\ref{quartic_r}) vanishes, then the angle dependence of $p_\pm$, $p_{\pm,\text{neg}}$ is absent and, furthermore, $p_\pm=-p_{\pm,\text{neg}}$ (see also Sec.~\ref{sec::roots}). In this case, angle integration in Eq.~(\ref{M_def}) is trivial, while integration over the absolute value $p$ of momentum can be replaced with an integration over $p^2$. For such integrals, we can extend the integration contour to~$-\infty$ and close it through the upper half-plane. Then the value of the integral is given by a sum of residues at the $p^2_{\pm}$ poles of Eqs.~(\ref{zero_r_poles}) that acquire finite imaginary parts due to a $\varepsilon\to\varepsilon+i/2\tau$ shift.

Hence, computation of the matrix $\mathcal M$ is straightforward when $\alpha_{\text{\tiny R}}=0$, $\Delta_{\text{sd}}=0$, or $\bb n=\bb n_{\perp}$. In this section, we calculate $\xi_0$, $\xi_\parallel$, and $\xi_\perp$, for the first and third cases. In the next section, we use the first two cases as reference points for perturbative analysis of these functions.


\subsection{Vanishing spin-orbit coupling}
\label{sec::no_SOC}
We will study the case of $\alpha_{\text{\tiny R}}=0$ first. In the absence of SOC, conservation of spin brings a technical difficulty to the calculation of~$\mathcal T$. Namely, at zero frequency and zero momentum, the matrix of disorder-averaged advanced-retarded spin-spin correlators $\mathcal M(I-\mathcal M)^{-1}$ that enters Eq.~(\ref{T_matrix_def}) cannot be finite. Indeed, using the formulas of Appendix~\ref{sec::structure_of_M} with $\alpha_{\text{\tiny R}}=0$, one finds
\be
\label{MvsU_no_SOC}
\mathcal M=I-\frac{2\varsigma\tau\Delta_{\text{sd}}}{1+(2\tau\Delta_{\text{sd}})^2}U(I-2\varsigma\tau\Delta_{\text{sd}} U),
\e
so that $I-\mathcal M$ is proportional to $U$. But $\det U=0$ and, therefore, $\mathcal M(I-\mathcal M)^{-1}=\infty$. Physically, this divergence is caused by the absence of linear response of electron spins polarized along $\bb n$ to time-dependent homogeneous perturbations of $J_{\text{sd}}$ (cf. Sec. 8.3 in Ref.~\onlinecite{Rammer}). Nevertheless, even in the limit of zero momentum and zero frequency, STT, GD, and ESR remain finite, since the series
\be
\label{T_no_SOC_sum}
\mathcal T=U\mathcal M+U\mathcal M^2+U\mathcal M^3+\dots
\e
actually converges.

The sum in Eq.~(\ref{T_no_SOC_sum}) is most easily calculated in the diagonal representation of $U$,
\be
\label{Udiag_def}
U=V U_{\text{diag}} V^\dagger, \qquad U_{\mathrm{diag}}= \diag{(i,-i,0)},
\e
which is defined by the unitary matrix
\be
V=
\begin{pmatrix}
\frac{i\,n_y-n_x n_z}{\sqrt{2(n_x^2+n_y^2)}} &  -\frac{i\, n_y+n_x n_z}{\sqrt{2(n_x^2+n_y^2)}} & n_x \\
-\frac{i\, n_x+n_y n_z}{\sqrt{2(n_x^2+n_y^2)}} &  \frac{i\, n_x-n_y n_z}{\sqrt{2(n_x^2+n_y^2)}} & n_y \\
\frac{\sqrt{n_x^2+n_y^2}}{\sqrt{2}} & \frac{\sqrt{n_x^2+n_y^2}}{\sqrt{2}} & n_z \\
\end{pmatrix}.
\e
Introducing $\mathcal M_U = V^\dagger\mathcal M V$ and making use of the relation $U_{\text{diag}}=U_{\text{diag}}P$, to take care of the potential divergence, we can rewrite Eq.~(\ref{T_no_SOC_sum}) as
\be
\label{T_no_SOC_sum_2}
\mathcal T=V U_{\text{diag}}(P\mathcal M_U+P\mathcal M_U^2+P\mathcal M_U^3+\dots) V^\dagger,
\e
where, according to Eqs.~(\ref{MvsU_no_SOC})~and~(\ref{Udiag_def}),
\be
P\mathcal M_U^k=
\diag{\left(\left[1+2i\varsigma\tau\Delta_{\text{sd}}\right]^{-k},
\left[1-2i\varsigma\tau\Delta_{\text{sd}}\right]^{-k},
0\right)}.
\e
Summation in Eq.~(\ref{T_no_SOC_sum_2}) is trivially performed, leading to
\begin{multline}
\mathcal T= -\frac{\varsigma}{2\tau\Delta_{\text{sd}}}V U_{\text{diag}}^2 V^\dagger=-\frac{\varsigma}{2\tau\Delta_{\text{sd}}}U^2=\\
\frac{\varsigma}{2\tau\Delta_{\text{sd}}}\left(I-\mathcal R^T \mathcal R\right)=\widetilde{\mathcal T}-\widetilde{X}\mathcal R,
\end{multline}
where $\widetilde{\mathcal T}=(\varsigma/2\tau\Delta_{\text{sd}})I$ represents the gauge of Eq.~(\ref{definiton_of_T_tilded}) and we have used the first identity of Eq.~(\ref{gauge_identity_1}).

The above result clearly corresponds to 
$t_0=\varsigma/2\tau\Delta_{\text{sd}}$ and $t_\parallel=t_\perp=0$, or
\be
\xi_0=\frac{\varsigma\Delta_{\text{sd}} A m}{2\pi\hbar^2 S}, \qquad \xi_\parallel=\xi_\perp=0.
\e
Hence, Gilbert damping and the nonadiabatic spin-transfer torque are both absent when $\alpha_{\text{\tiny R}}=0$, as it should be in the model with no SOC, spin-dependent disorder, or other sources of spin relaxation.

The parameter $\xi_0$ defines the effective spin renormalization (due to conduction electrons) in the LLG equation as~\cite{Tatara2008review} $\xi_0=-\delta S_{\text{eff}}/S$. In fact, for $\alpha_{\text{\tiny R}}=0$, the effective spin renormalization coincides with actual spin renormalization. Indeed, without SOC, all electrons are polarized along~$\pm\bb n$, and, for the calculation of the total electron spin in a unit cell,
\begin{multline}
\delta S=\delta S_\uparrow-\delta S_\downarrow=\frac{\varsigma}{2}\left(N_+-N_-\right)=
\\
\frac{\varsigma A}{8\pi^2\hbar^2}\left[\int\limits_{\varepsilon_+(\bb p)\leq\varepsilon}{p\,dp d\phi_{\bb p}}-\int\limits_{\varepsilon_-(\bb p)\leq\varepsilon}{p\,dp d\phi_{\bb p}}\right],
\end{multline}
one may use $\varepsilon_{\pm}(\bb p)\leq\varepsilon\Leftrightarrow p^2\leq 2m(\varepsilon\mp\Delta_{\text{sd}})$ to obtain
\be
\label{spin_renormalization}
\delta S=-\frac{\varsigma\Delta_{\text{sd}}A m}{2\pi\hbar^2}.
\e
Thus, $\delta S=-\xi_0 S=\delta S_{\text{eff}}$ in this case.

In Appendix~\ref{sec::spin_susceptibility}, we compute spin susceptibility of the system for $\alpha_{\text{\tiny R}}\neq 0$ and demonstrate that the spin renormalization does not depend on the SOC strength. At the same time, the effective spin renormalization does. Moreover, the identity $\delta S_{\text{eff}}=\delta S$ is, in fact, a very specific case. It holds either for vanishing spin-orbit interaction, or at some particular value of $\Delta_{\text{so}}\approx\Delta_{\text{sd}}$, as one can learn from Table~\ref{table} and Fig.~\ref{fig::plots} (we recall that $\Delta_{\text{so}}=\vert\alpha_{\text{\tiny R}}\vert\sqrt{2m\varepsilon}$ characterizes the SOC-induced splitting of the spectral branches).


\subsection{Perpendicular-to-the-plane magnetization}
\label{sec::out_of_plane}
Now we turn to the $\bb n=\bb n_\perp$ regime. The formulas of Appendix~\ref{sec::structure_of_M} are not applicable in this case. Nevertheless, one can perform the integration in Eq.~(\ref{M_def}) directly, utilizing the expression for the Green's function of Eq.~(\ref{g_in_momentum_space}) with $\sin{\theta}=0$ (and $\bb n=\bb e_z\cos{\theta}$). It follows that
\begin{multline}
\mathcal M=\left[1+4\tau^2(\Delta_{\text{sd}}^2+\Delta_{\text{so}}^2)\right]^{-1}
\Bigl(\left[1+2(\tau\Delta_{\text{so}})^2\right] P
\\
+\left[1+4(\tau\Delta_{\text{sd}})^2\right](I-P)-2\varsigma\tau\Delta_{\text{sd}}P\,UP\Bigr)
\end{multline}
and, after some arithmetic,
\begin{multline}
\label{T__nperp}
\mathcal T=\frac{\varsigma}{2\tau\Delta_{\text{sd}}}
\left[1-\frac{\left(\tau\Delta_{\text{so}}^2\right)^2}{\Delta_{\text{sd}}^2+\tau^2(2\Delta_{\text{sd}}^2+\Delta_{\text{so}}^2)^2}\right]P
\\
+\frac{1}{2}\left[
\frac{\Delta_{\text{so}}^2\left[1+2\tau^2(2\Delta_{\text{sd}}^2+\Delta_{\text{so}}^2)\right]}{\Delta_{\text{sd}}^2+\tau^2(2\Delta_{\text{sd}}^2+\Delta_{\text{so}}^2)^2}\right]P\,UP.
\end{multline}

Substitution of this result into Eqs.~(\ref{STTandGD_def_new}) shows that, in this case, both $\bb T^{\text{STT}}$ and $\bb T^{\text{GD}}$ are represented as linear combinations of two vector forms: $\partial\bb n_\parallel$ and $\bb n_\perp \times\partial\bb n_\parallel$. Since $\bb n=\bb n_\perp$ and, thus, $\partial\bb n_{\perp}=0$, the coefficients in front of these forms should be recognized as $t_0$ and $t_\parallel$, respectively. With the help of Eq.~(\ref{spin_renormalization}), we, therefore, find
\begin{subequations}
\begin{gather}
\xi_0=-\frac{\delta S}{S}\left[1-\frac{\left(\tau\Delta_{\text{so}}^2\right)^2}{\Delta_{\text{sd}}^2+\tau^2(2\Delta_{\text{sd}}^2+\Delta_{\text{so}}^2)^2}\right],
\\
\label{xi_par_nperp}
\xi_\parallel=\Bigl\vert\frac{\delta S}{S}\Bigr\vert\tau\Delta_{\text{sd}}\left[\frac{\Delta_{\text{so}}^2\left[1+2\tau^2(2\Delta_{\text{sd}}^2+\Delta_{\text{so}}^2)\right]}{\Delta_{\text{sd}}^2+\tau^2(2\Delta_{\text{sd}}^2+\Delta_{\text{so}}^2)^2}\right].
\end{gather}
\end{subequations}

For a fixed $\bb n=\bb n_\perp$, however, one cannot directly define $\xi_{\perp}$. Indeed, the latter function, in this case, is a prefactor in front of the vanishing vector form \mbox{$\bb n \times\partial\bb n_\perp$} and, in principle, can be even taken arbitrary. The only way to assign a clear meaning to $\xi_{\perp}$, here, is to consider its asymptotic behaviour at small values of~$\sin{\theta}$. Namely, one should expand the integrands in Eq.~(\ref{M_def}) up to $\sin^2{\theta}$ and, after the integration, compute the coefficients of the decomposition of Eq.~(\ref{Tsymm}) with the same accuracy. Application of a $\sin{\theta}\to 0$ limit in Eq.~(\ref{t_perp_def}), afterwards, will lead to
\be
\label{xi_perp_nperp}
\xi_\perp=\Bigl\vert\frac{\delta S}{S}\Bigr\vert\tau\Delta_{\text{sd}}
\left[\frac{1}{2}\frac{\Delta_{\text{so}}^2\left[1+(2\tau\Delta_{\text{sd}})^2\right]}{\Delta_{\text{sd}}^2+\tau^2(2\Delta_{\text{sd}}^2+\Delta_{\text{so}}^2)^2}\right].
\e

One may use Eqs.~(\ref{xi_par_nperp}) and (\ref{xi_perp_nperp}) to evaluate the strength of the rotational anisotropy of GD and the nonadiabatic STT, given $\bb n\approx\bb n_\perp$. We see, for example, that, for small $\sin{\theta}$, the ratio
\be
\label{rotat_anysot_nperp}
\xi_\parallel/\xi_\perp =2+ \frac{\Delta_{\text{so}}^2}{\Delta_{\text{sd}}^2+(1/2\tau)^2}+\mathcal O(\sin^2{\theta}),
\e
exceeds 2, making the rotational anisotropy considerable even if SOC is weak. At the same time, for strong spin-orbit coupling, $\xi_\parallel$ can potentially be orders of magnitude larger than $\xi_\perp$ (see also Fig.~\ref{fig::plots}).

For the perpendicular-to-the-plane magnetization, GD was analyzed previously in Ref.~[\onlinecite{Garate2009GD}] under an additional assumption of large chemical potential. Our result for the Gilbert damping coefficient $\xi_\parallel$, given by Eq.~(\ref{xi_par_nperp}), coincides with the expression on the right hand side of Eq.~(25) of Ref.~[\onlinecite{Garate2009GD}], to an overall factor that we were unable to identify (most likely, it is equal to 4). The $\tau\to\infty$ limit of the same expression was derived recently in Ref.~[\onlinecite{baglai}] (with another overall factor). This paper also mentions the role of the diagonal terms of the GD tensor on ESR.

A separate study of the nonadiabatic STT (also limited to the $\bb n=\bb n_\perp$ case) was reported in Ref.~[\onlinecite{Garate2009STT}]. As we have shown above, this torque should be fully determined by the very same function $\xi_\parallel$ as is GD. The authors, however, ignored vertex corrections, and, as it seems, overlooked this fact. In any case, their results differ from those of Eq.~(\ref{xi_par_nperp}).

\section{Asymptotic expansions}
\label{sec::asymptotics}
We proceed with a calculation of the $\xi_i$ expansions in either small $\alpha_{\text{\tiny R}}$ or small $\Delta_{\text{sd}}$. To perform such calculation, one should expand the integrands in Eq.~(\ref{M_def}) or, alternatively, in Eqs.~(\ref{gamma_coeff_gen}), with respect to the corresponding variable. Then the result can be integrated over the poles, provided by Eqs.~(\ref{zero_SOC_poles})~and~(\ref{zero_J_poles}), respectively (where $\varepsilon$ should be replaced with $\varepsilon+i/2\tau$).

\begingroup
\begin{table*}
\begin{center}
\begin{tabular}{c|c|c|c}
&$\xi_0/(-\frac{\delta S}{S})$ or $\delta S_{\text{eff}}/\delta S$\phantom{$\Bigr|$} &
$\xi_{\parallel}/(\vert\frac{\delta S}{S}\vert\tau\Delta_{\text{sd}})$& 
$\xi_{\perp}/(\vert\frac{\delta S}{S}\vert\tau\Delta_{\text{sd}})$
\\\hline
$\alpha_{\text{\tiny R}}=0$ & $1$ & 0 & 0
\\\hline
$\mathcal{O}(\Delta_{\text{so}}^2)$
& 
$1+\dfrac{2(\tau\Delta_{\text{so}})^2}{1+(2\tau\Delta_{\text{sd}})^2}\dfrac{1-n_z^2}{1+n_z^2}$
&
$\dfrac{(\Delta_{\text{so}}/\Delta_{\text{sd}})^2}{1+(2\tau\Delta_{\text{sd}} )^2}\left[(2\tau\Delta_{\text{sd}})^2+\dfrac{2}{1+n_z^2}\right]$
& 
$\dfrac{(\Delta_{\text{so}}/\Delta_{\text{sd}})^2}{1+(2\tau\Delta_{\text{sd}} )^2}\dfrac{1+(2n_z\tau\Delta_{\text{sd}})^2}{1+n_z^2}$ \phantom{$\Biggr|$}
\\\hline
$\Delta_{\text{sd}}\to 0$
&
$\left(\dfrac{\Delta_{\text{sd}}}{\Delta_{\text{so}}}\right)^2\left[4n_z^2+\dfrac{1+n_z^2}{2\left(\tau\Delta_{\text{so}}\right)^2}\right]$
& 
$2+\dfrac{1}{\left(\tau\Delta_{\text{so}}\right)^2}$
& 
$\dfrac{1}{2\left(\tau\Delta_{\text{so}}\right)^2}$ \phantom{$\Biggr|$}
\\\hline
$\bb n=\bb n_{\perp}$
&
$1-\dfrac{(\tau\Delta_{\text{so}}^2)^2}{\Delta_{\text{sd}}^2+\tau^2(2\Delta_{\text{sd}}^2+\Delta_{\text{so}}^2)^2}$ 
& 
$\dfrac{\Delta_{\text{so}}^2\left[1+2\tau^2(2\Delta_{\text{sd}}^2+\Delta_{\text{so}}^2)\right]}{\Delta_{\text{sd}}^2+\tau^2(2\Delta_{\text{sd}}^2+\Delta_{\text{so}}^2)^2}$\phantom{$\Biggr|$}
& 
$\dfrac{1}{2}\dfrac{\Delta_{\text{so}}^2\left[1+(2\tau\Delta_{\text{sd}})^2\right]}{\Delta_{\text{sd}}^2+\tau^2(2\Delta_{\text{sd}}^2+\Delta_{\text{so}}^2)^2}$ \phantom{$\biggr|$}
\end{tabular}
\end{center}
\caption{\label{table}Closed-form results and asymptotic expansions for the dimensionless functions $\xi_0$, $\xi_{\parallel}$, and $\xi_{\perp}$ that define anisotropic spin-transfer torques, Gilbert damping, and effective spin renormalization. The results are expressed in terms of the energy scales $\Delta_{\text{sd}}=\vert J_{\text{sd}}\vert S$ and $\Delta_{\text{so}}=\vert\alpha_{\text{\tiny R}}\vert\sqrt{2\varepsilon m}$ that describe, respectively, the exchange and spin-orbit-induced splitting. The second row shows the expansion up to the second order in $\Delta_{\text{so}}$. The third row provides the leading order terms of the expansion with respect to small $\Delta_{\text{sd}}$. Spin renormalization is defined in Eq.~(\ref{spin_renormalization}) by $\delta S=-J_{\text{sd}}S A m/2\pi\hbar^2$.}
\end{table*}
\endgroup

\subsection{Weak spin-orbit coupling}
\label{sec::weak_SOC}
Keeping the notation of Sec.~\ref{sec::no_SOC} for the matrices $\mathcal M$ and $\mathcal T$ in the absence of SOC, below we use the symbols $\delta\mathcal M$ and $\delta\mathcal T$ to represent the respective contributions provided by finite $\alpha_{\text{\tiny R}}$.

Since $\delta\mathcal M\neq 0$, the result of matrix inversion in
\be
\mathcal T+\delta\mathcal T=U(\mathcal M+\delta\mathcal M)(I-\mathcal M-\delta\mathcal M)^{-1}
\e
is finite, making the analysis straightforward yet rather cumbersome. Retaining only proportional to $\alpha^2_{\text{\tiny R}}$ terms in $\delta\mathcal M$ (see Appendix~\ref{sec::weak_SOC_extra} for explicit formulas), we obtain
\be
\delta\mathcal T=\delta c_2 P+\delta c_3 U+\delta c_4 U^2+\dots,
\e
where dots represent terms that do not contribute to the $\delta\widetilde{\mathcal T}$ gauge in the $\alpha^2_{\text{\tiny R}}$ order and
\begin{subequations}
\begin{gather}
\delta c_2=\frac{\Delta_{\text{so}}^2}{2\Delta_{\text{sd}}^2}\frac{1}{1+n_z^2},\\
\delta c_3=-\frac{\tau\Delta_{\text{so}}^2}{\varsigma\Delta_{\text{sd}}\left[1+\left(2\tau\Delta_{\text{sd}}\right)^2\right]}\frac{1-n_z^2}{1+n_z^2},\\
\delta c_4=-\frac{\Delta_{\text{so}}^2}{2\Delta_{\text{sd}}^2}\frac{1+\left(2n_z\tau\Delta_{\text{sd}}\right)^2}{1+\left(2\tau\Delta_{\text{sd}}\right)^2}\frac{1}{1+n_z^2}.
\end{gather}
\end{subequations}
Then, utilizing Eqs.~(\ref{t_i_def}) with $c_i$ replaced by $\delta c_i$, we arrive at the second-order expansions in small SOC strength for the functions $\xi_i$. Those are collected in the second row of Table~\ref{table}.

We may again use the obtained results to quantify the rotational anisotropy of GD and the nonadiabatic STT by computing the ratio
\be
\label{rotat_anysot_smallSOC}
\xi_\parallel/\xi_\perp=
2 + \frac{1-n_z^2}{n_z^2+1/(2\tau\Delta_{\text{sd}})^2}+\mathcal O(\Delta_{\text{so}}^2).
\e
For weak spin-orbit coupling, the rotational anisotropy is minimal when magnetization is perpendicular to the plane and increases for the magnetization approaching the in-plane direction.

\begin{figure*}[t]
\includegraphics[width=\textwidth]{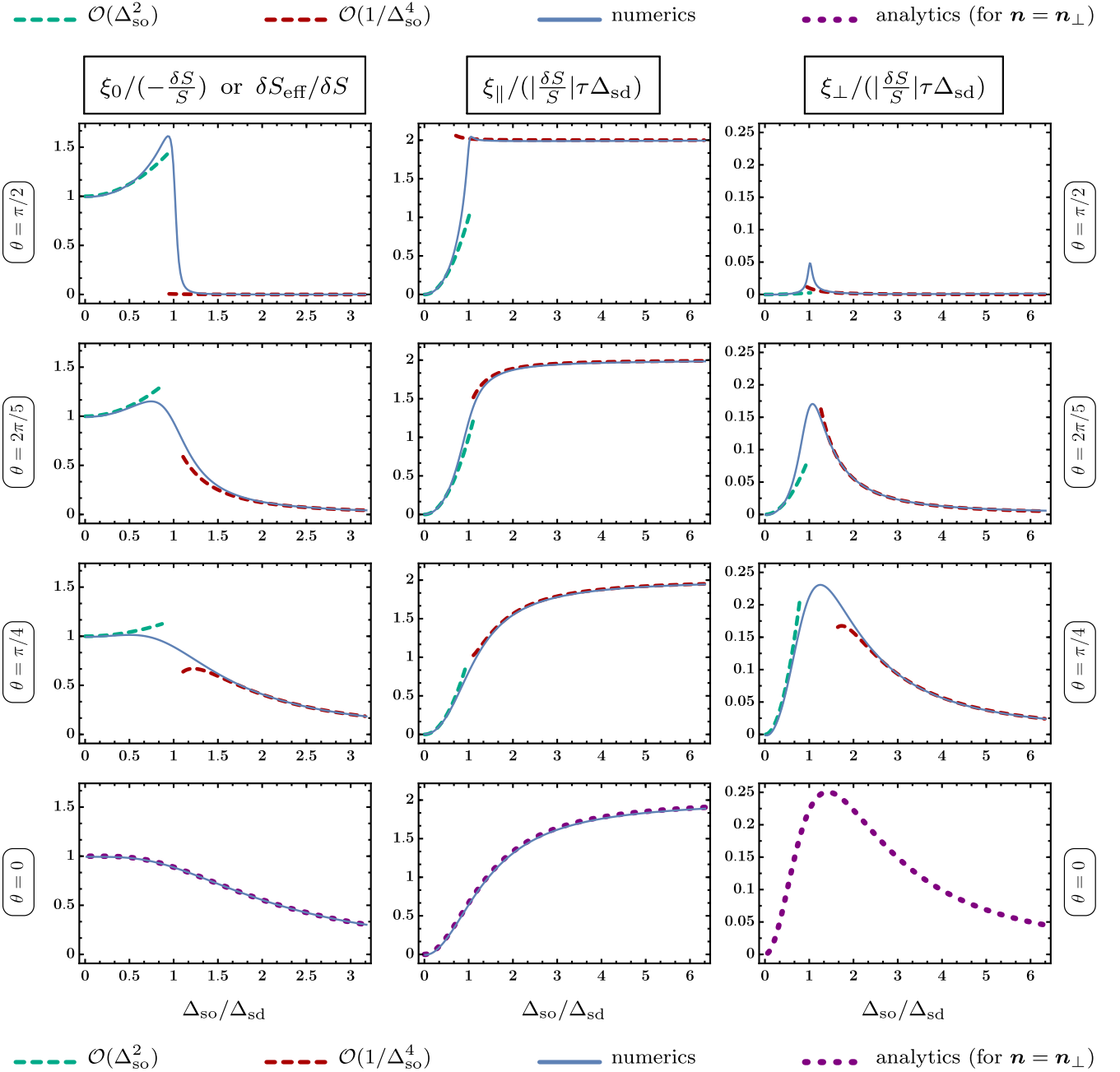}
\caption{Dimensionless functions $\xi_0$, $\xi_{\parallel}$, and $\xi_{\perp}$ that define anisotropic spin-transfer torques, Gilbert damping, and effective spin renormalization as functions of the spin-orbit coupling strength $\alpha_{\text{\tiny R}}$ for four different polar angles of magnetization ($n_z=\cos{\theta}$). The notations coincide with those of Table~\ref{table}. We use the dimensionless combinations $\varepsilon\tau=50$, $\tau\Delta_{\text{sd}}=10$. Since for $\theta=0$ it is impossible to compute $\xi_{\perp}$ numerically, only analytical result is shown. The $\mathcal O\left(1/\Delta_{\text{so}}^4\right)$ expansion is addressed in Appendix~\ref{sec::next_order}.}
\label{fig::plots}
\end{figure*}

We also note that the asymptotic expansions up to the order $\alpha_{\text{\tiny R}}^2$ allow us to estimate the orientational anisotropy of $\xi_i$. Employing the notation $\xi_i=\xi_i(n_z^2)$, we find
\begin{subequations}
\begin{align}
\xi_0(0)-\xi_0(1)&=\frac{2(\tau\Delta_{\text{so}})^2}{1+(2\tau\Delta_{\text{sd}})^2},\\
\xi_\parallel(0)-\xi_\parallel(1)&=\frac{1}{1+(2\tau\Delta_{\text{sd}})^2}\,\frac{\Delta_{\text{so}}^2}{\Delta_{\text{sd}}^2},\\
\label{perp_orient_anysot_smallSOC}
\xi_\perp(0)-\xi_\perp(1)&=\frac{1-(2\tau\Delta_{\text{sd}})^2}{1+(2\tau\Delta_{\text{sd}})^2}\,\frac{\Delta_{\text{so}}^2}{2\Delta_{\text{sd}}^2},
\end{align}
\end{subequations}
for weak SOC. Clearly, $\xi_0$ and $\xi_\parallel$ are both maximal for $\bb n_\perp=0$. On the other hand, the expression on the right hand side of Eq.~(\ref{perp_orient_anysot_smallSOC}) can change sign, depending on the value of $\tau\Delta_{\text{sd}}$. Therefore, the orientational anisotopy of $\xi_\perp$ in a ``clean'' system ($\tau\Delta_{\text{sd}}\gg 1$) differs from that in a ``dirty'' one (Fig.~\ref{fig::plots} corresponds to the case of a ``clean'' system).

Interestingly, at $\alpha_{\text{\tiny R}}=0$ the matrix function $\delta\mathcal T$ turns out to be discontinuous. Namely, its elements have finite limits for $\alpha_{\text{\tiny R}}\to 0$. This discontinuity has, however, no physical consequences, since the matrix $\delta\mathcal T$ itself is not gauge invariant. In the $\delta\widetilde{\mathcal T}$ gauge, the discontinuity is removed and, thus, it does not affect the physically relevant quantities $\xi_0$, $\xi_{\parallel}$, and $\xi_{\perp}$. This property demonstrates the importance of full analysis of all components of the STT and GD tensors.

\subsection{Weak exchange interaction}
\label{sec::weak_sd}
Up to the linear order in $\Delta_{\text{sd}}$, we have
\be
\mathcal M=
\frac{I+2(\tau\Delta_{\text{so}})^2 P}{1+4(\tau\Delta_{\text{so}})^2}
-2\varsigma\tau\Delta_{\text{sd}}\frac{U+4(\tau\Delta_{\text{so}})^2 P\,UP}{\left[1+4(\tau\Delta_{\text{so}})^2\right]^2}.
\e
This corresponds to the following coefficients of the decomposition of Eq.~(\ref{Tsymm}),
\begin{subequations}
\begin{gather}
c_1=\frac{1}{4(\tau\Delta_{\text{so}})^2}, \quad
c_2=1+\frac{1}{4(\tau\Delta_{\text{so}})^2}, \quad
\\
c_3=-\frac{\varsigma\tau\Delta_{\text{sd}}}{4(\tau\Delta_{\text{so}})^4}, \quad
c_4=0, \quad
\\
c_5=-\frac{\varsigma\tau\Delta_{\text{sd}}\left[1+8(\tau\Delta_{\text{so}})^2\right]}{4(\tau\Delta_{\text{so}})^4}, \quad
c_6=0.
\end{gather}
\end{subequations}
Substituting the latter expressions into Eqs.~(\ref{t_i_def}), one obtains the leading-order contributions to $\xi_i$ in the limit of small $\Delta_{\text{sd}}$. The respective results are presented in the third row of Table~\ref{table}. Using them, we can find yet another expression for the ratio
\be
\label{rotat_anysot_smallexc}
\xi_\parallel/\xi_\perp= 2 + (2\tau\Delta_{\text{so}})^2+\mathcal{O}(\Delta_{\text{sd}}^2).
\e

Remarkably, the rotational anisotropy of GD and the nonadiabatic STT, $\xi_\parallel/\xi_\perp= 2$, persists to both limits
\be
\Delta_{\text{sd}}\ll \Delta_{\text{so}}\ll 1/\tau
\quad\,\,\text{and}\quad\,\,
\Delta_{\text{so}}\ll \Delta_{\text{sd}}\ll 1/\tau,
\e
in which the Fermi surfaces defined in Eq.~(\ref{spectrum}) are not only essentially isotropic but, at the same time, do get strongly broadened by the disorder (the broadening $1/\tau$ exceeds the splitting of the subbands).

It is also interesting to mention that, for small values of $\Delta_{\text{sd}}$, the nonadiabatic spin-transfer torque dominates over the adiabatic one: $\xi_{\parallel,\perp}/\xi_0\propto 1/\Delta_{\text{sd}}$. This agrees with the intuitive logic that, for a weak exchange between conduction and localized spins, the former would rather not adiabatically follow the direction of the latter.


\section{Discussion}
\label{sec::discussion}
\subsection{Role of vertex corrections}
We would like to begin this final section by stressing that it is 
the accurate consideration of vertex corrections that is responsible for the established vector structures of anisotropic STT, GD, and ESR, as well as for the relation between them. Practically none of this would be seen from an uncontrolled analysis that ignores vertex corrections.

For example, if one does not apply the disorder dressing to the current vertex $\bb v$, the relation of Eq.~(\ref{STT_to_T}) will no longer be valid. Instead, the STT tensor, in this case, will contain 18 additional nonzero components of different symmetries, which one might by mistake interpret as physical torques.


\subsection{Renormalization of spin}
In Sec.~\ref{sec::no_SOC}, we have demonstrated that, in the limit of vanishing SOC, the ESR factor $\delta S_{\text{eff}}=-\xi_0 S$ does coincide with the actual total electron
spin in a unit cell $\delta S=-J_{\text{sd}}S A m/2\pi\hbar^2$. On the other hand, this equality breaks down for finite~$\alpha_{\text{\tiny R}}$, and the ratio $\delta S_{\text{eff}}/\delta S$ starts to depend on all of the parameters of the system, including scattering time (see Table~\ref{table} and Fig.~\ref{fig::plots}).

For large values of spin-orbit-induced splitting $\Delta_{\text{so}}$, the quantity $\xi_0$ (which determines ESR) understandably decays due to the effective randomization of the electron spin direction induced by SOC. What is, however, rather interesting, is that, for relatively small values of~$\alpha_{\text{\tiny R}}$, the ESR factor $\delta S_{\text{eff}}$ exceeds $\delta S$, reaching the maximum value at $\Delta_{\text{so}}\approx\Delta_{\text{sd}}$. We do not have an intuitive explanation for such behaviour.


\subsection{LLG equation}
It is instructive to compare the microscopic LLG Eq.~(\ref{LLG_short}) to its conventional phenomenological counterpart. In the absence of spin-orbit, thermal, and other torques that we do not consider in this study, the latter equation reads 
\begin{align}
\label{LLG_conventional}
\partial_t\bb n=\gamma \bb n \times \bb H_{\text{eff}}&+\left(\bb j_{\text{s}}\cdot\bb\nabla \right) \bb n \nonumber
\\
-\alpha[\bb n &\times \partial_t \bb n]-\beta[\bb n \times\left(\bb j_s\cdot\bb\nabla\right) \bb n],
\end{align}
where the vector quantity $\bb j_{\text{s}}$ is interpreted as the phenomenological spin-polarized current, while the parameters $\alpha$ and $\beta$ define Gilbert damping and the nonadiabatic spin-transfer torque, respectively. The latter is also commonly referred to as the $\beta$-torque. The adiabatic spin-transfer torque is represented by the term $\lt(\bb j_{\text{s}}\cdot\bb\nabla \rt) \bb n$, while $\bb H_{\text{eff}}$ stands for effective field contributions.

First, taking into account Eqs.~(\ref{STT_and_GD_general}), we can rewrite the microscopic LLG Eq.~(\ref{LLG_short}) in a form which is similar to that of Eq.~(\ref{LLG_conventional}),
\begin{align}
\label{LLG_for_experimentalists}
\partial_t\bb n=\bar\gamma \bb n \times \bb H_{\text{eff}}&+\left(\bb j_{\text{s}}\cdot\bb\nabla \right) \bb n \nonumber
\\
-\alpha_{\parallel\phantom{\perp}\nphantom{\parallel}}[\bb n &\times \partial_t \bb n_{\parallel\phantom{\perp}\nphantom{\parallel}}]
-\beta_{\parallel\phantom{\perp}\nphantom{\parallel}}[\bb n \times\left(\bb j_s\cdot\bb\nabla\right) \bb n_{\parallel\phantom{\perp}\nphantom{\parallel}}] \nonumber
\\
-\alpha_\perp[\bb n &\times \partial_t \bb n_\perp]-\beta_\perp[\bb n \times\left(\bb j_s\cdot\bb\nabla\right) \bb n_\perp],
\end{align}
where
\begin{subequations}
\begin{gather}
\label{spin_current}
\bb j_{\text{s}}=\bb v_{\text{d}}\frac{\xi_0}{1-\xi_0}=-\bb v_{\text{d}}\frac{\delta S_{\text{eff}}}{S+\delta S_{\text{eff}}},
\\
\label{alpha_beta_gamma_def}
\alpha_{\parallel,\perp}=\frac{\xi_{\parallel,\perp}}{1-\xi_0}, \quad \beta_{\parallel,\perp}=\frac{\xi_{\parallel,\perp}}{\xi_0}, \quad \bar\gamma=\frac{\gamma}{1-\xi_0}
\end{gather}
\end{subequations}
and each of the quantities $\bb j_{\text{s}}$, $\alpha_{\parallel,\perp}$, $\beta_{\parallel,\perp}$, $\bar\gamma$ depend on the orientation of the vector $\bb n$. For the particular 2D Rashba FM model system considered in this paper,
\begin{subequations}
\begin{alignat}{2}
\bb j_{\text{s}}&=\bb j_{\text{s}}(n_z^2),& \qquad \alpha_{\parallel,\perp}&=\alpha_{\parallel,\perp}(n_z^2),\\
\beta_{\parallel,\perp}&=\beta_{\parallel,\perp}(n_z^2),& \qquad \bar\gamma&=\bar\gamma(n_z^2).
\end{alignat}
\end{subequations}
We see that the microscopic LLG Eq.~(\ref{LLG_for_experimentalists}) is essentially anisotropic, in contrast with the phenomenological LLG Eq.~(\ref{LLG_conventional}). Namely, the coefficients $\alpha$ and $\beta$ got split into two components each. Moreover, the new coefficients $\alpha_{\parallel,\perp}$ and $\beta_{\parallel,\perp}$ as well as the other parameters of the LLG equation became dependent on the direction of magnetization. We note that the splitting of the GD coefficient~$\alpha$ has been reported, for a Rashba FM, in Ref.~[\onlinecite{Phenomenological_GD}].

Next, let us comment on the microscopic definiton of the spin-polarized current formulated in Eq.~(\ref{spin_current}). Normally, if spins of conduction electrons (travelling with the characteristic velocity $\bb v$) adiabatically follow the direction of $\bb n$, one assumes $\bb j_{\text{s}}=-\bb v\,\delta S/(S+\delta S)$, where $\delta S$ is a contribution from conduction electrons to the total spin of the system. In this case, Eq.~(\ref{LLG_conventional}) can be simply viewed as a manifestation of the total angular momentum conservation (for $\bb n \times \bb H_{\text{eff}}=0$),
\be
(S+\delta S)\partial_t\bb n+\delta S\left(\bb v\cdot\bb\nabla \right) \bb n=0.
\e
where $-\delta S\left(\bb v\cdot\bb\nabla \right)\bb n$ is the rate of angular momentum transfer from conduction to total spin.

The definition of the vector quantity $\bb j_{\text{s}}$, given by Eq.~(\ref{spin_current}), provides a perfect generalization of the above logic for a system with finite Rashba SOC. Indeed, conduction spins no longer follow the direction of $\bb n$ (due to, e.g., nonzero damping). Nevertheless, $-\delta S_{\text{eff}}\left(\bb v_{\text{d}}\cdot\bb\nabla \right)\bb n$ still has a meaning of the rate of ``angular momentum transfer'' from the effective conduction spin $\delta S_{\text{eff}}$ to the total $S+\delta S_{\text{eff}}$. Importantly, it was a fully controllable accurate microscopic treatment of the problem that led us to Eq.~(\ref{spin_current}). (We identified the drift velocity $\bb v_{\text{d}}$ as a ``proportionality coefficient'' between the STT and GD tensors and observed that 
the adiabatic spin-transfer torque and ESR are described by the same quantity $\xi_0$.)

Finally, for the sake of historical integrity, let us also mention that the equalities $\alpha_\parallel=\beta_\parallel$ and $\alpha_\perp=\beta_\perp$, in this system, are equivalent~\cite{last_resort_comment} to the relation
\be
\label{unphysical}
\delta S_{\text{eff}}=-S/2,
\e
which appears to be rather unphysical.


\subsection{Material derivative and moving reference frame}
In the presence of the anisotropic STT and GD of Eqs.~({\ref{STT_and_GD_general}}), it is natural to analyse the microscopic LLG Eq.~(\ref{LLG_short}) in such a frame, where the effect of the nonadiabatic spin-transfer torque is absent. Namely, in the frame that moves with the classical drift velocity of conduction electrons $\bb v_{\text{d}}$. One may use a nice analogy to continuum mechanics as an illustration of this fact.

Indeed, despite the essentially anisotropic character of both $\bb T^{\text{STT}}$ and $\bb T^{\text{GD}}$, their sum is conveniently expressed in the LLG Eq.~(\ref{LLG_short}) via the operator of material derivative $D_t=\partial_t+\left(\bb v_\text{d}\cdot\bb\nabla \right)$ as
\begin{multline}
\label{LLG_fluid}
(1-\xi_0)D_t\bb n=\gamma \bb n \times \bb{H}_{\text{eff}}+\left(\bb v_\text{d}\cdot\bb\nabla \right)\bb n
-\xi_{\parallel}\left[\bb n\times D_t\bb n_{\parallel}\right]
\\
-\xi_{\perp}\left[\bb n \times D_t\bb{n}_{\perp}\right]+\dots,
\end{multline}
where we have moved the term $\xi_0 D_t\bb n$ to the left hand side and added $\left(\bb v_\text{d}\cdot\bb\nabla \right)\bb n$ to both sides. By considering conduction electrons as a ``fluid'' flowing with the drift velocity $\bb v_{\text{d}}$, one may interpret the material derivatives of Eq.~(\ref{LLG_fluid}) as the change rates of components of $\bb n$ that are associated with the electronic ``fluid parcels''. Thus, in the moving (``flowing'') frame, $\bb r'=\bb r-\bb v_d t$, the material derivatives $D_t$ are automatically replaced~\cite{reference_frame_comment} by the ordinary time derivatives $\partial_t$.

In other words, in the moving reference frame, Eq.~(\ref{LLG_fluid}) takes the form of the LLG equation
\begin{multline}
(1-\xi_0)\partial_t\bb n=\gamma \bb n \times \bb{H}_{\text{eff}}+\left(\bb v_\text{d}\cdot\bb\nabla \right)\bb n
-\xi_{\parallel}\left[\bb n\times \partial_t\bb n_{\parallel}\right]
\\
-\xi_{\perp}\left[\bb n \times \partial_t\bb{n}_{\perp}\right]+\dots
\end{multline}
that comprises the analogue of the adiabatic torque $\left(\bb v_\text{d}\cdot\bb\nabla \right)\bb n$, two components of damping, and (represented here by dots) all other possible torques. As long as the latter are absent, the dynamics of a magnetic texture, governed by such equation (under mediate currents and in the absence of magnetic field), is likely to be a motion with zero terminal velocity (as it is~\cite{1stLiZhang2004, Thiaville2005}, in the isotropic case, for domain walls). For a general situation, current-induced magnetic dynamics can differ significantly. Nevertheless, it should still be more convenient to perform the analysis once the effect of the nonadiabatic STT has been accounted for by switching to the ``flowing'' frame.

Interestingly, any ``propagating'' texture of the form $\bb n(\bb r,t)\hspace{-0.2ex}=\hspace{-0.2ex}\bb \zeta(\bb r-\bb v_\text{d}t)\hspace{-0.2ex}=\hspace{-0.2ex}\bb\zeta_{\bb r}(t)$ nullifies the sum $\bb T^{\text{STT}}+\bb T^{\text{GD}}$. Hence, for such textures, the LLG Eq.~(\ref{LLG_short}) reads
\be
\label{LLG_wavy}
d\bb \zeta_{\bb r}/dt=\gamma \bb\zeta_{\bb r} \times \bb{H}_{\text{eff}}+\dots,
\e
where $\bb r$ can be regarded as a parameter. If one takes into account only spin-transfer torques and fieldlike spin-orbit torque, solutions of this equation will have an oscillatory character. Note that Eq.~(\ref{LLG_wavy}) is different from the LLG equation
\be
0=\gamma \bb\zeta_{\bb r} \times \bb{H}_{\text{eff}}
\e
that describes the uniform motion of the ground state in the presence of the Galilean invariance [the case $\alpha=\beta$ in Eq.~(\ref{LLG_conventional})]~\cite{Barnes2005alphabeta, DuinevsKeldysh2007,Tserkovnyak2008review, Tatara2008review}.


\subsection{Response to electric current}
So far, we have computed spin-transfer torques as a linear response of the system to the external electric field $\bb E$. In experiment, however, it is not the electric field but rather the electric current $\bb j$ which is externally applied. To relate spin torques to the latter, one should compute the conductivity tensor~$\hat\sigma$ and, afterwards, use the identity
\be
\bb E=\hat\sigma^{-1}\bb j
\e
to replace $\bb E$ with $\bb j$. Importantly, the conductivity tensor has to be computed up to the linear order in first magnetization gradients $\nabla_\alpha n_\beta$.


\subsection{Relation to Edelstein effect}
\label{Edelstein_section}
It is worth noting that some of our results can be independently benchmarked.  As it was suggested in Ref.~\onlinecite{Stiles2013DMIvsA}, there exists a connection between some particular pairs of quantities in the model of Eq.~(\ref{Hamiltonian}), as, e.g., between the Dzyaloshinskii-Moriya interaction strength and the exchange stiffness, or between spin-orbit torques and spin-transfer torques. The latter relation is relevant to our study. 

A general interpretation of the approach described in Ref.~\onlinecite{Stiles2013DMIvsA} would be the following. Suppose there exists a quantity $\mathcal F(\alpha_{\text{\tiny R}})$ which, for the model with $\alpha_{\text{\tiny R}}=0$, depends on the gradients of $\bb n$, such that 
\be
\mathcal F(0)=F(\nabla_x \bb n, \nabla_y \bb n).
\e
Then, up to the linear order with respect to $\alpha_{\text{\tiny R}}$, one would obtain~\cite{alpha_different_sign}
\be
\label{theorem}
\mathcal F(\alpha_{\text{\tiny R}})=\mathcal F(0)+\alpha_{\text{\tiny R}}\left[\frac{\partial}{\partial\alpha_{\text{\tiny R}}} F(\widetilde{\nabla}_x \bb n, \widetilde{\nabla}_y \bb n)\right]_{\alpha_{\text{\tiny R}}=0},
\e
where
\be
\widetilde{\nabla}_i \bb n=\nabla_i \bb n + \frac{2m\alpha_{\text{\tiny R}}}{\hbar}[\bb n\times[\bb e_z\times \bb e_i]].
\e

Let us now choose three functions $\mathcal F_i(\alpha_{\text{\tiny R}})$ to be the components of the vector $\bb T^{\text{STT}}$. Using the expression for the quantity $\xi_0$ in the limit $\alpha_{\text{\tiny R}}=0$ (see Table~\ref{table}), we can write
\be
\bb T^{\text{STT}}=\frac{eA}{2\pi\hbar}J_{\text{sd}}\tau(\bb E\cdot\bb\nabla)\bb n.
\e
From Eq.~(\ref{theorem}) we, then, find another contribution to the generalized torque in the $\propto\alpha_{\text{\tiny R}}$ order
\be
\label{SOT_from_STT}
\bb T^{\text{SOT}}=\frac{2m\alpha_{\text{\tiny R}}}{\hbar}\frac{eA}{2\pi\hbar}J_{\text{sd}}\tau[\bb n\times[\bb e_z\times \bb E]],
\e
which is precisely the expression for the Edelstein effect~\cite{Edelstein1990} in a form of a fieldlike torque on magnetization. In a similar way, vanishing of the functions $\xi_\parallel$ and $\xi_\perp$ when $\alpha_{\text{\tiny R}}=0$ can be translated into the absence~\cite{AdoSOT2017} of the antidamping SOT in the model of Eq.~(\ref{Hamiltonian}).

The result of Eq.~(\ref{SOT_from_STT}) coincides with the direct derivation of SOT, for the model of Eq.~(\ref{Hamiltonian}), that has been reported previously~\cite{AdoSOT2017}. A more compact and accurate form of this derivation is also presented in Appendix \ref{sec::VCtoV}. Such independent consistency check adds to the credibility of our results.

\section*{Conclusions}
We have presented a thorough microscopic analysis of STT, GD, and ESR, for the particular 2D FM system with Rashba spin-orbit coupling and spin-independent Gaussian white-noise disorder. Assuming arbitrary direction of magnetization, we have established the exact relation between these effects. We have introduced the notion of the matrix gauge transformation for magnetization-dependent phenomena and used it to express spin-transfer torques, Gilbert damping, and effective spin renormalization in terms of meaningful vector forms. The latter allowed us to quantify the SOC-induced anisotropy of the former. We have analysed, both analytically and numerically, three dimensionless functions that fully define anisotropic STT, GD, and ESR. We have also generalized the concept of spin-polarized current, computed spin susceptibility of the system, and obtained a number of other results.

It would be an interesting challenge to observe the anisotropy of STT experimentally. It might be possible to do this by measuring current-induced corrections to the magnon spectrum asymmetry that is normally associated with the Dzyaloshinskii-Moriya interaction. We also believe that, to some extent, the anisotropy of STT and GD might explain the differences in dynamics of domain walls (and skyrmions) with different characteristics.

\acknowledgments

We would like to thank Jairo Sinova for pointing out a number of flaws in the original version of the manuscript. We are also grateful to Artem Abanov, Arne Brataas, Sergey Brener, Ivan Dmitriev, Rembert Duine, Olena Gomonay, Andrew Kent, Alessandro Principi, Alireza Qaiumzadeh, and Yaroslav Tserkovnyak for helpful discussions. This research was supported by the JTC-FLAGERA Project GRANSPORT and by the Dutch Science Foundation NWO/FOM 13PR3118. M.T. acknowledges the support from the Russian Science Foundation under Project 17-12-01359. 


\appendix

\section{Vertex corrections to velocity operator; spin-orbit torque}
\label{sec::VCtoV}
In order to compute vertex corrections to the velocity operator $\bb v=\bb p/m-\alpha_{\text{\tiny R}}[\bb e_z\times\bb\sigma]$, we first apply a single impurity line to the scalar part of the latter,
\be
\label{p_one_dressing_def}
\left(\bb p/m\right)^{1\times \text{dr}}=\frac{1}{m\tau}\int{
\frac{d^2 p}{(2\pi)^2}\,
g^R\left(\bb p/m\right) g^A}.
\e
Due to the fact that the momentum operator $\bb p$ commutes with the Green's functions $g^{R,A}$, the above relation can be equivalently written as
\be
\label{p_one_dressing_further}
\left(\bb p/m\right)^{1\times \text{dr}}=\frac{i}{m}\int{
\frac{d^2 p}{(2\pi)^2}\,
\left(\bb p/m\right)\left(g^R-g^A\right)},
\e
where we have used the Hilbert's identity of Eq.~(\ref{Hilbert}).

The subsequent analysis follows the route of Sec.~\ref{sec::self-energy}. Integration over the absolute value of momentum in Eq.~(\ref{p_one_dressing_further}) is performed by computing residues at $p=p_\pm$. Symmetrization of the obtained result, with respect to the transformation~\cite{angle_integration} $\varphi\to\pi-\varphi$, leads to
\begin{multline}
\left(\bb p/m\right)^{1\times \text{dr}}=\int_0^{2\pi}\frac{d\varphi}{2\pi}
\Bigl(
\alpha_{\text{\tiny R}}\left(1+r W_4\right)[\bb e_z\times\bb\sigma]
\\+
\left(\alpha_{\text{\tiny R}}+r W_5\right)\left\{\bb n_\parallel [\bb\sigma\times\bb n]_z -\left(\bb n_\parallel\cdot\bb\sigma\right)[\bb e_z\times\bb n]\right\}\cos{2\varphi}
\\+
(W_6+W_7\,\bb n\cdot\bb\sigma)[\bb e_z\times\bb n]\sin{\varphi}
\Bigr),
\end{multline}
where $W_i=W_i\left(r^2,u\,(r^2)\right)$ are some functions of the parameter $r^2$ and $\varphi$-independent parameters of the model. Again, all terms that contain $W_i$ vanish identically after integration over the angle and we conclude that
\be
\label{p_one_dressing}
\left(\bb p/m\right)^{1\times \text{dr}}=\alpha_{\text{\tiny R}}[\bb e_z\times\bb\sigma].
\e

Next, we observe that the corrected by an impurity ladder velocity operator $\bb v^{\text{vc}}$ can be recast in the form
\begin{multline}
\bb v^{\text{vc}}=\bigl\{\bb p/m-\alpha_{\text{\tiny R}}[\bb e_z\times\bb\sigma]\bigr\}^{\text{vc}}
=\\
\bb p/m+\bigl\{\left(\bb p/m\right)^{1\times \text{dr}}-\alpha_{\text{\tiny R}}[\bb e_z\times\bb\sigma]\bigr\}^{\text{vc}}.
\end{multline}
According to Eq.~(\ref{p_one_dressing}), expression inside the brackets on the second line vanishes, leading us to the desired result,
\be
\label{Vvc_again}
\bb v^{\text{vc}}=\bb p/m,
\e
which coincides with Eq.~(\ref{VCtoV}) of the main text. Note that, since the momentum operator commutes with the Green's functions, Eq.~(\ref{Vvc_again}) determines both advanced-retarded and retarded-advanced vertex corrections to the velocity operator.

One immediate consequence of Eqs.~(\ref{p_one_dressing})~and~(\ref{Vvc_again}) is a trivial form of spin-orbit torque in the considered interface Rashba model. Indeed, it was conjectured in Ref.~\onlinecite{AdoSOT2017} that the antidamping SOT, in this model, is identically absent, while the field-like SOT is entirely isotropic. To prove the conjecture, we use the Kubo formula for SOT 
\be
\bb T^{\text{SOT}}=
\frac{e J_{\text{sd}} A}{2\pi\hbar^2}
\int{
\frac{d^2 p}{(2\pi)^2}\,
\mtr{\left\{\hat{\bb T}\, g^R\left(\bb v^{\text{vc}}\cdot\bb E\right)g^A\right\}}}.
\e
Substituting $\bb v^{\text{vc}}=\bb p/m$ and using Eq.~(\ref{p_one_dressing_def}), we immediately find 
\be
\bb T^{\text{SOT}}=
\frac{e J_{\text{sd}} A m\tau}{2\pi\hbar^2}
\mtr{\left\{\hat{\bb T}\left(\left(\bb p/m\right)^{1\times \text{dr}}\cdot\bb E\right)\right\}},
\e
Finally, with the help of Eqs.~(\ref{T_operator})~and~(\ref{p_one_dressing}), we obtain the expression for spin-orbit torque, 
\begin{multline}
\bb T^{\text{SOT}}=
\frac{e J_{\text{sd}} A m\tau\alpha_{\text{\tiny R}}}{2\pi\hbar^2}
\mtr{\bigl\{[\bb\sigma\times\bb n]\left([\bb e_z\times\bb\sigma]\cdot\bb E\right)\bigr\}}
=\\
\frac{e J_{\text{sd}} A m\tau\alpha_{\text{\tiny R}}}{\pi\hbar^2}[\bb n\times[\bb e_z\times \bb E]],
\end{multline}
which coincides with that of Eq.~(\ref{SOT_from_STT}), as expected.


\section{Vanishing of $\delta\mathcal{T}^{\text{STT}}$}
\label{sec::deltaT_vanishes}
We will now prove that the absence of the spin component in the vertex corrected velocity operator $\bb v^{\text{vc}}$ nullifies the contribution $\delta\mathcal{T}^{\text{STT}}$ to the STT tensor of Eq.~(\ref{delta_STT_tensor}). Using cyclic permutations under the matrix trace and the fact that $\bb v^{\text{vc}}=\bb p/m$ commutes with any function of momentum, one can rewrite Eq.~(\ref{delta_STT_tensor}) as
\be
\delta\mathcal{T}^{\text{STT}}_{\alpha\beta\gamma\delta}=
-\frac{e\Delta_{\text{sd}}^2 A}{2\pi\hbar S}
\int{
\frac{d^2 p}{(2\pi)^2}\frac{p_{\beta}\tau}{2m}\mtr{\left[\Lambda_1+\Lambda_2\right]
}}
\e
with
\begin{subequations}
\begin{gather}
\label{lambda_1_def}
\Lambda_1=\Bigl(
v_{\gamma}\,g^A\,\hat{T}^{\text{vc}}_\alpha\,g^R\,\sigma_{\delta} -
\sigma_{\delta}\,g^A\,\hat{T}^{\text{vc}}_\alpha\,g^R\,v_{\gamma}
\Bigr)\frac{g^R g^A}{i\tau},
\\
\label{lambda_2_def}
\begin{aligned}
\Lambda_2=\Bigl(&
\sigma_{\delta}\,g^A\,v_{\gamma}\,g^A\,\hat{T}^{\text{vc}}_\alpha -
v_{\gamma}\,g^A\,\sigma_{\delta}\,g^A\,\hat{T}^{\text{vc}}_\alpha
\Bigr)\frac{g^R g^A}{i\tau}
\\
-\Bigl(&
\hat{T}^{\text{vc}}_\alpha\,g^R\,v_{\gamma}\,g^R\,\sigma_{\delta} -
\hat{T}^{\text{vc}}_\alpha\,g^R\,\sigma_{\delta}\,g^R\,v_{\gamma}
\Bigr)\frac{g^R g^A}{i\tau}.
\end{aligned}
\end{gather}
\end{subequations}
In Eq.~(\ref{lambda_1_def}), we employ the Hilbert's indentity of Eq.~(\ref{Hilbert}) to replace the factor $g^R g^A/i\tau$ with $g^R-g^A$ and again use cyclic permutations to obtain
\be
\begin{aligned}
\Lambda_1{}={}&
\hat{T}^{\text{vc}}_\alpha\,g^R\,\sigma_{\delta}\,g^R\,v_{\gamma}\,g^A -
\hat{T}^{\text{vc}}_\alpha\,g^R\,v_{\gamma}\,g^R\,\sigma_{\delta}\,g^A
\\
{}-{}&
\hat{T}^{\text{vc}}_\alpha\,g^R\,\sigma_{\delta}\,g^A\,v_{\gamma}\,g^A +
\hat{T}^{\text{vc}}_\alpha\,g^R\,v_{\gamma}\,g^A\,\sigma_{\delta}\,g^A.
\end{aligned}
\e
A similar procedure is performed to simplify the expression for $\Lambda_2$. We note, however, that terms with only retarded or only advanced Green's functions, in Eq.~(\ref{lambda_2_def}), should be disregarded~\cite{comment_on_Kubo}. Hence, $g^R g^A/i\tau$ is replaced with $g^R$ in the first line of Eq.~(\ref{lambda_2_def}) and with $-g^A$ in the second line. After moving the torque operator to the first place in each term,
\be
\begin{aligned}
\Lambda_2{}={}&
\hat{T}^{\text{vc}}_\alpha\,g^R\,\sigma_{\delta}\,g^A\,v_{\gamma}\,g^A -
\hat{T}^{\text{vc}}_\alpha\,g^R\,v_{\gamma}\,g^A\,\sigma_{\delta}\,g^A
\\
{}+{}&
\hat{T}^{\text{vc}}_\alpha\,g^R\,v_{\gamma}\,g^R\,\sigma_{\delta}\,g^A -
\hat{T}^{\text{vc}}_\alpha\,g^R\,\sigma_{\delta}\,g^R\,v_{\gamma}\,g^A,
\end{aligned}
\e
we conclude that $\Lambda_1+\Lambda_2=0$ and, therefore, $\delta\mathcal{T}^{\text{STT}}=0$ as well.


\section{Structure of $\mathcal M$}
\label{sec::structure_of_M}
Using Green's function of Eq.~(\ref{g_in_momentum_space}) we compute the matrix trace in Eq.~(\ref{M_def}) and further symmetrize the integrands with respect to the transformation~\cite{angle_integration} $\varphi\to\pi-\varphi$. This results in the decomposition
\be
\mathcal M=\gamma_1 I+\gamma_2 P+\gamma_3 U+\gamma_4 U^2+\gamma_5 P\, U P+\gamma_6 P\, U^2 P
\e
where the coefficients are given in the integral form,
\begin{subequations}
\label{gamma_coeff_gen}
\begin{gather}
\gamma_1=2\left[
\left(\Delta_{\text{sd}}^2+\left\vert\varepsilon+i/2\tau\right\vert^2\right)\mathcal I_0-2\left(\varepsilon+\delta_{\text{so}}\right)\mathcal I_1+\mathcal I_2
\right],
\\
\gamma_2=-4\left[
\frac{2\delta_{\text{so}}n_z^2}{1-n_z^2}\mathcal I_1+
\frac{\left(1+n_z^2\right)\varsigma\Delta_{\text{sd}}}{\sqrt{1-n_z^2}}\mathcal J_1-
\frac{1+n_z^2}{1-n_z^2}\mathcal J_2
\right],
\\
\gamma_3=-\frac{2}{\tau}\left[
\varsigma\Delta_{\text{sd}}\mathcal I_0-\frac{1}{\sqrt{1-n_z^2}}\mathcal J_1
\right],
\\
\gamma_4=4\varsigma\Delta_{\text{sd}}\left[
\varsigma\Delta_{\text{sd}}\mathcal I_0-\frac{1}{\sqrt{1-n_z^2}}\mathcal J_1
\right],
\\
\gamma_5=-
\frac{2}{\tau\sqrt{1-n_z^2}}\mathcal J_1,
\\
\gamma_6=-4\left[
\frac{2\delta_{\text{so}}}{1-n_z^2}\mathcal I_1+
\frac{\varsigma\Delta_{\text{sd}}}{\sqrt{1-n_z^2}}\mathcal J_1-
\frac{2}{1-n_z^2}\mathcal J_2
\right],
\end{gather}
\end{subequations}
with $\delta_{\text{so}}=m \alpha^2_{\text{\tiny R}}$ and
\begin{subequations}
\label{gamma_coeff_gen_int}
\begin{gather}
\mathcal I_k=\int{
\frac{d^2 p}{(2\pi)^2}\,\frac{\left(2m\tau\right)^{-1}\left(p^2/2m\right)^k}{\vert\varepsilon-\varepsilon_+(\bb p)+i/2\tau\vert^2\vert\varepsilon-\varepsilon_-(\bb p)+i/2\tau\vert^2}},
\\
\mathcal J_k=\int{
\frac{d^2 p}{(2\pi)^2}\,\frac{\left(2m\tau\right)^{-1}\left(\alpha_{\text{\tiny R}}p\sin{\varphi}\right)^k}{\vert\varepsilon-\varepsilon_+(\bb p)+i/2\tau\vert^2\vert\varepsilon-\varepsilon_-(\bb p)+i/2\tau\vert^2}}.
\end{gather}
\end{subequations}
Some of Eqs.~(\ref{gamma_coeff_gen}) formally become invalid when $\bb n=\bb n_\perp$. However, structure of $\mathcal M$ and $\mathcal T$ in the respective case was analysed directly in Sec.~\ref{sec::out_of_plane}.


\section{Structure of $\mathcal M^k$}
\label{sec::structure_of_M^k}
We have already demonstrated that
\be
\mathcal M\in\Span{\mathcal L},\,\,\,\,\,\mathcal L=\{I,\, P,\, U,\, U^2,\, P\,UP,\, P\,U^2P\},
\e
Let us now prove that any natural power of $\mathcal M$ belongs to the same linear span,
\be
\mathcal M^{k}\in\Span{\mathcal L},\qquad \forall k\in\mathbb N.
\e

The operation of matrix product, by itself, is not closed on $\Span{\mathcal L}$. Moreover, 14 of 36 elements of $\mathcal L\times\mathcal L$ do not belong to $\Span{\mathcal L}$. On the other hand, a combination of two such elements (matrices $P\,U$ and $UP$),
\be
P\,U+UP=\{P,U\}=U+P\,UP,
\e
obviously does. Similarly, the remaining 12 ``unsuitable'' elements of $\mathcal L\times\mathcal L$ do form 6 pairs, such that the corresponding anticommutators (namely, $\{P,U^2\}$, $\{P\,UP,U\}$, $\{P\,U^2P,U\}$, $\{P\,UP,U^2\}$, $\{P\,U^2P,U^2\}$, and $\{P\,UP,P\,U^2P\}$) belong to $\Span{\mathcal L}$. 

In general, the following statement holds: operation of matrix anticommutation sends elements of $\mathcal L\times\mathcal L$ to a linear span of $\mathcal L$,
\be
\label{anticommutator_pre_binary}
\{\phantom{\cdot},\phantom{\cdot}\}:\mathcal L\times\mathcal L\to \Span{\mathcal L}.
\e
Taking into account the fact that anticommutator is a bilinear map, we deduce from Eq.~(\ref{anticommutator_pre_binary}):
\be
\label{anticommutator_binary}
\{\phantom{\cdot},\phantom{\cdot}\}:\Span{\mathcal L}\times\Span{\mathcal L}\to \Span{\mathcal L}.
\e
Finally, since for arbitrary $k$ we have
\be
\mathcal M^k=\frac{1}{2}\{\mathcal M,\mathcal M^{k-1}\},
\e
the desired result, $\mathcal M^k\in\Span{\mathcal L}$, is proven by induction. 


\section{Spin susceptibility in the presence of SOC}
\label{sec::spin_susceptibility}
In this Appendix, the total spin $\delta\bb{S}$ of conduction electrons in a unit cell of the area $A$ is computed for a general case of $\alpha_{\text{\tiny R}}\neq 0$. We use the following standard definition:
\be
\label{conduction_spin_def}
\delta\bb{S}=\frac{A}{2\pi i}\int{d \epsilon\,f(\epsilon)}\int{\frac{d^2 p}{(2\pi\hbar)^2}\mtr{\left[\frac{\bb
\sigma}{2}\left(G^A-G^R\right)\right]}},
\e
where $f$ stands for the Fermi-Dirac distribution,
\be
f(\epsilon)=\left(1+\exp{\left[(\epsilon-\varepsilon)/T\right]}\right)^{-1},
\e
and $G^{A,R}$ refers to the momentum-dependent Green's function of Eq.~(\ref{G_def}). We will first consider the in-plane component of $\delta\bb S$.

Matrix trace calculation followed by an integration over $\epsilon$, in Eq.~(\ref{conduction_spin_def}), gives
\begin{subequations}
\label{S_parallel_traced}
\begin{gather}
\delta S_x=A\int{\frac{d^2 p}{(2\pi\hbar)^2}\,
\frac{\varsigma\Delta_{\text{sd}} n_x-\alpha_{\text{\tiny R}}p_y}{\varepsilon_+(\bb p)-\varepsilon_-(\bb p)}(f_+-f_-)},\\
\delta S_y=A\int{\frac{d^2 p}{(2\pi\hbar)^2}\,
\frac{\varsigma\Delta_{\text{sd}} n_y+\alpha_{\text{\tiny R}}p_x}{\varepsilon_+(\bb p)-\varepsilon_-(\bb p)}(f_+-f_-)},
\end{gather}
\end{subequations}
where $f_\pm=f(\varepsilon_\pm(\bb p))$. It is convenient to introduce the quantity $\delta S_+=\delta S_x+i \delta S_y$. For the latter, we find
\begin{multline}
\label{delta_S_plus_1}
\delta S_+=\frac{A}{4\alpha_{\text{\tiny R}}}\int{\frac{d^2 p}{(2\pi\hbar)^2}\,
(f_+-f_-)}\\
\times\left(i\frac{\partial}{\partial p_x}-\frac{\partial}{\partial p_y}\right)
\left[\varepsilon_+(\bb p)-\varepsilon_-(\bb p)\right],
\end{multline}
where we took advantage of the fact that the fractions in Eqs.~(\ref{S_parallel_traced}) can be expressed as the derivatives with respect to the components of momentum. In the zero-temperature limit, one can use Green's theorem to reduce the double integrals in Eq.~(\ref{delta_S_plus_1}) to the integrals over the closed curves $C_\pm=\{\bb p\mid\varepsilon_\pm(\bb p)=\varepsilon\}$,
\begin{subequations}
\begin{gather}
\delta S_+=\delta S_+^+ + \delta S_+^-,
\\
\delta S_+^{\pm}=\pm\frac{A}{4\alpha_{\text{\tiny R}}}\int\limits_{C_{\pm}}{\frac{d p_x+i d p_y}{(2\pi\hbar)^2}\left[\varepsilon_+(\bb p)-\varepsilon_-(\bb p)\right]}.
\end{gather}
\end{subequations}

Next, we follow the approach used by K.-W. Kim \textit{et~al.} in Ref.~\onlinecite{KimMATH2016}. Using the variable $w=p_x+i p_y$ and the relation $\varepsilon_\pm(\bb p)=p^2/2m\pm[\varepsilon_+(\bb p)-\varepsilon_-(\bb p)]/2$, we find
\be
\label{S_pm_w}
\delta S_+^{\pm}=\frac{A}{16\pi^2\hbar^2\alpha_{\text{\tiny R}}}\int\limits_{C_{\pm}}{dw\left(2\varepsilon-\frac{w^*w}{m}\right)},
\e
where $w^* w=p^2$ and $C_\pm=\{w\mid\varepsilon_\pm(w,w^*)=\varepsilon\}$ are now regarded as contours in the complex $w$-plane. Since the contours are closed, Eq.~(\ref{S_pm_w}) is further simplifed to
\be
\label{S_pm_w_2}
\delta S_+^{\pm}=-\frac{A}{16\pi^2\hbar^2 m\alpha_{\text{\tiny R}}}\int\limits_{C_{\pm}}{dw\,w^*w}.
\e

In order to perform integration in Eq.~(\ref{S_pm_w_2}), we solve the equation $\varepsilon_\pm(w,w^*)=\varepsilon$ for $w^*$ and express the result as a function of $w\in C_\pm$,
\be
\label{w_conjugate_solved}
w^*
=\frac{2m}{w^2}\left(
w\left[\varepsilon+m\alpha_{\text{\tiny R}}^2\right]-i m\alpha_{\text{\tiny R}}\varsigma\Delta_{\text{sd}}n_+\pm\sqrt{R}
\right),
\e
where $n_+=n_x+i n_y$ and $R$ is a cubic function of $w$. Different signs in front of the square root in Eq.~(\ref{w_conjugate_solved}) correspond to two different functions $w^*=w^*_\pm(w)$ of $w\in C_\pm$, respectively. We do not specify which sign corresponds to which function. Such ambiguity, however, does not affect the final result for $\delta S_+$. Indeed, it can be proven~\cite{KimMATH2016} that all three zeroes of $R$ are of the form $w_k=i r_k n_+$ with real $r_k$. Then, from the general relation
\begin{multline}
\left[\varepsilon-\varepsilon_+(w,w^*)\right]
\left[\varepsilon-\varepsilon_-(w,w^*)\right]
=-R\\
+\left(\frac{w^*w}{2m}-\left[\varepsilon+m\alpha_{\text{\tiny R}}^2\right]
+\frac{i m\alpha_{\text{\tiny R}}\varsigma\Delta_{\text{sd}} n_+}{w}\right)^2,
\end{multline}
we learn that
\be
\left[\varepsilon-\varepsilon_+(w_k,w^*_k)\right]
\left[\varepsilon-\varepsilon_-(w_k,w^*_k)\right]\geq 0
\e
and, thus, $\varepsilon_-(w_k,w^*_k)<\varepsilon\Rightarrow\varepsilon_+(w_k,w^*_k)\leq\varepsilon$. Hence, all the singularities of $w^*_-$ that lie inside the contour $C_-$ are, in fact, located inside or, at most, on the contour $C_+$ (note that $C_+$ is inside $C_-$). Disregarding the case~\cite{comment_on_singularities} $w_k\in C_\pm$ and using Cauchy integral theorem, we can shrink~\cite{comment_on_shrinking} $C_-$ in Eq.~(\ref{S_pm_w_2}) to obtain
\be
\delta S_+=-\frac{A}{16\pi^2\hbar^2 m\alpha_{\text{\tiny R}}}\int\limits_{C_{+}}
{dw\,\left(w_+^* + w_-^*\right)w},
\e
so that the terms $\pm\sqrt{R}$, in Eq.~(\ref{w_conjugate_solved}), do not contribute to $\delta S_+$. The only remaining singularity of the integrand is located at the origin and, by the residue theorem,
\be
\label{S_parallel}
\delta S_+
=-\frac{\varsigma\Delta_{\text{sd}}Am}{2\pi\hbar^2}n_+
\,\,\,
\,\,\text{or}\,\,
\,\,\,
\delta \bb S_\parallel
=-\frac{\varsigma\Delta_{\text{sd}}Am}{2\pi\hbar^2}\bb n_\parallel,
\e
which completes the computation of the in-plane component of $\delta\bb S$.

In order to calculate $\delta S_z$, it is useful to introduce the ``magnetization'' vector $\bb M=\varsigma \Delta_{\text{sd}}\bb n$. In terms \mbox{of $\bb M$,} one can straightforwardly establish the ``thermodynamic'' relation $\delta S_i=\partial \Omega/\partial M_i$, where $\Omega$ has a meaning of the electronic grand potential in a unit cell,
\begin{subequations}
\begin{gather}
\Omega=-T\frac{A}{2\pi i}\int{d \epsilon\,g(\epsilon)}\int{\frac{d^2 p}{(2\pi\hbar)^2}\mtr{\left[G^A-G^R\right]}},
\\
g(\epsilon)=\log{\left(1+\exp{\left[(\varepsilon-\epsilon)/T\right]}\right)}.
\end{gather}
\end{subequations}
We further note that, according to Eq.~(\ref{S_parallel}), $\delta S_x$ and $\delta S_y$ do not depend on $M_z$. Therefore, equating the second derivatives, we find 
\be
\frac{\partial\delta S_z}{\partial M_\alpha}=\frac{\partial^2\Omega}{\partial M_\alpha\partial M_z}=\frac{\partial\delta S_\alpha}{\partial M_z}=0,
\e
where $\alpha=x, y$. As a result, $\delta S_z$ does not depend on $M_x$ and $M_y$ and, thus, can be computed for $M_x=M_y=0$ (or, equivalently, for $n_x=n_y=0$).

From Eq.~(\ref{conduction_spin_def}) we obtain
\be
\label{S_perpendicular_traced}
\delta S_z=A\int{\frac{d^2 p}{(2\pi\hbar)^2}\,
\frac{\varsigma\Delta_{\text{sd}} n_z}{\varepsilon_+(\bb p)-\varepsilon_-(\bb p)}(f_+-f_-)},
\e
which, for $n_x=n_y=0$, can be integrated over the momentum angle with the result
\be
\label{S_perpendicular_integral}
\delta S_z=A\frac{\varsigma\Delta_{\text{sd}}n_z}{4\pi\hbar^2}\int\limits_0^\infty{p dp\,
\frac{f_+-f_-}{\sqrt{\Delta_{\text{sd}}^2+\left(\alpha_{\text{\tiny R}}p\right)^2}}}.
\e
At zero temperature, the integration domain in Eq.~(\ref{S_perpendicular_integral}) is reduced to a finite interval $p_+<p<p_-$, where $p_\pm$ are given by  Eq.~(\ref{nperp_poles}). After some algebraic practice, we finally arrive at
\be
\label{S_perpendicular}
\delta S_z=\left. A\frac{\varsigma\Delta_{\text{sd}}n_z}{4\pi\hbar^2\alpha_{\text{\tiny R}}^2}
\sqrt{\Delta_{\text{sd}}^2+\left(\alpha_{\text{\tiny R}}p\right)^2}\right\vert_{p_+}^{p_-}=-\frac{\varsigma\Delta_{\text{sd}}Am}{2\pi\hbar^2}n_z.
\e

Combining the results of Eqs.~(\ref{S_parallel})~and~(\ref{S_perpendicular}) into 
a single vector form
\be
\label{S_full}
\delta \bb S
=-\frac{\varsigma\Delta_{\text{sd}}Am}{2\pi\hbar^2}\bb n,
\e
we see that, on average, even for finite values of spin-orbit coupling strength $\alpha_{\text{\tiny R}}$, spins of conduction electrons, in the equilibrium, are aligned with the local magnetization. Moreover, the spin susceptibility tensor is fully isotropic and is expressed by a single scalar parameter
\be
\delta S=-\left\vert\delta \bb S\right\vert=-\frac{\varsigma\Delta_{\text{sd}}Am}{2\pi\hbar^2},
\e
which coincides with that given by Eq.~(\ref{spin_renormalization}) of the main text.

\onecolumngrid

\section{Expansion of $\mathcal M$ up to $\alpha^2_{\text{\tiny R}}$}
\label{sec::weak_SOC_extra}
Expansion of Eqs.~(\ref{gamma_coeff_gen}) up to $\alpha^2_{\text{\tiny R}}=\Delta_{\text{so}}^2/2\varepsilon m$ provides us with the coefficients
\begin{subequations}
\begin{align}
\delta\gamma_1=-&\left[\frac{2\tau\Delta_{\text{so}}}{1+(2\tau\Delta_{\text{sd}})^2}\right]^2\left[1+(2 n_z \tau\Delta_{\text{sd}})^2\right],&
\quad
\delta\gamma_2=2&\left[\frac{\tau\Delta_{\text{so}}}{1+(2\tau\Delta_{\text{sd}})^2}\right]^2\left[1-(1+2n_z^2)(2\tau\Delta_{\text{sd}})^2\right],
\\
\delta\gamma_3=&\left[\frac{4\tau\Delta_{\text{so}}}{1+(2\tau\Delta_{\text{sd}})^2}\right]^2\frac{1+(2 n_z \tau\Delta_{\text{sd}})^2}{1+(2\tau\Delta_{\text{sd}})^2}\varsigma\tau\Delta_{\text{sd}},&
\quad
\delta\gamma_4=-2&\left[\frac{4\tau^2\Delta_{\text{so}}\Delta_{\text{sd}}}{1+(2\tau\Delta_{\text{sd}})^2}\right]^2\frac{1+(2 n_z \tau\Delta_{\text{sd}})^2}{1+(2\tau\Delta_{\text{sd}})^2},
\\
\delta\gamma_5=-2&\left[\frac{2\tau\Delta_{\text{so}}}{1+(2\tau\Delta_{\text{sd}})^2}\right]^2 \varsigma\tau\Delta_{\text{sd}},&
\quad
\delta\gamma_6=-&\left[\frac{4\tau^2\Delta_{\text{so}}\Delta_{\text{sd}}}{1+(2\tau\Delta_{\text{sd}})^2}\right]^2
\end{align}
\end{subequations}
of the decomposition that we refer to in Sec.~\ref{sec::weak_SOC}:\,\,\,$\delta\mathcal M=\delta\gamma_1 I+\delta\gamma_2 P+\delta\gamma_3 U+\delta\gamma_4 U^2+\delta\gamma_5 P\, U P+\delta\gamma_6 P\, U^2 P.$


\section{$\mathcal O(1/\Delta_{\text{so}}^4)$ expansion of $\xi_i$ (limit of strong SOC)}
\label{sec::next_order}
The quantities $\xi_i$ are shown in the plots of Fig.~\ref{fig::plots} as functions of the spin-orbit coupling strength $\alpha_{\text{\tiny R}}$ (while keeping both $m$ and $\varepsilon$ constant). Therefore, the right ``tails'' of the curves can be properly fit using the asymptotic expansion with respect to the parameter $1/\Delta_{\text{so}}$. Such expansion can be obtained indirectly, from the expansion in small $\Delta_{\text{sd}}$. Below, for consistency with the results of Sec.~\ref{sec::weak_sd}, we list all the contributions to $\xi_i$ that do not exceed the fourth order in $1/\Delta_{\text{so}}$,
\begin{subequations}
\label{strong_SOC}
\begin{alignat}{4}
\label{0_strong_SOC}
\xi_0=-\frac{\delta S}{S}
& \Biggl[ &
& &
&\left(\frac{\Delta_{\text{sd}}}{\Delta_{\text{so}}}\right)^2\left[4n_z^2+\frac{1+n_z^2}{2\left(\tau\Delta_{\text{so}}\right)^2}\right]
&{}+{}& 6\left(\frac{\Delta_{\text{sd}}}{\Delta_{\text{so}}}\right)^4 \left[1-3n_z^2\right]n_z^2
\Biggr],
\\
\label{par_strong_SOC}
\xi_\parallel=\Bigl\vert\frac{\delta S}{S}\Bigr\vert\tau\Delta_{\text{sd}}
& \Biggl[ &
2+\frac{1}{\left(\tau\Delta_{\text{so}}\right)^2}& &
{}-{}& \left(\frac{\Delta_{\text{sd}}}{\Delta_{\text{so}}}\right)^2\left[4n_z^2-\frac{1-7n_z^2}{\phantom{2}\left(\tau\Delta_{\text{so}}\right)^2}\right]
&{}-{}& 4\left(\frac{\Delta_{\text{sd}}}{\Delta_{\text{so}}}\right)^4 \left[1-3n_z^2\right]n_z^2
\Biggr],
\\
\label{perp_strong_SOC}
\xi_\perp=\Bigl\vert\frac{\delta S}{S}\Bigr\vert\tau\Delta_{\text{sd}}
& \Biggl[ &
\frac{1}{2\left(\tau\Delta_{\text{so}}\right)^2}& &
{}+{}&\left(\frac{\Delta_{\text{sd}}}{\Delta_{\text{so}}}\right)^2\left[2n_z^2+\frac{1-5n_z^2}{2\left(\tau\Delta_{\text{so}}\right)^2}\right]
&{}+{}& 2\left(\frac{\Delta_{\text{sd}}}{\Delta_{\text{so}}}\right)^4 \left[1-5n_z^2\right]n_z^2
\Biggr].
\end{alignat}
\end{subequations}
Note that the expansion with respect to small~$\Delta_{\text{sd}}$ is different from the expansion with respect to large~$\Delta_{\text{so}}$.
\twocolumngrid


\bibliographystyle{apsrev4-1}
\bibliography{Bib}

\begin{thebibliography}{64}
\bibitem [{ref()}]{reference_frame_comment}%
  \BibitemOpen
  \href@noop {} {}\bibinfo {note} {That is -- by replacing $\bb n(\bb r, t)$
  with $\bb n(\bb r-\bb v_d t, t)$ in the LLG
  Eq.~(\ref{LLG_short}).}\BibitemShut {Stop}%
\bibitem [{com(0)}]{comment_on_hbar/2}%
  \BibitemOpen
  \href@noop {} {} \bibinfo {note} {Strictly speaking,
  it is the density divided by $\hbar/2$. We note that $\bb s(\bb r, t)$, here,
  is local quantum mechancial and disorder average of the operator
  $\bb\sigma$}\BibitemShut {NoStop}%
\bibitem [{ang()}]{angle_integration}%
  \BibitemOpen
  \href@noop {} {}\bibinfo {note} {Basically, we refer to the following
  procedure:
  $\int_0^\pi{f(\varphi)\,d\varphi}=\int_0^{\pi/2}{[f(\varphi)+f(\pi-\varphi)]\,d\varphi}=\int_0^\pi{\widetilde{f}(\varphi)}\,d\varphi$.
  The same applies to $\int_{\pi}^{2\pi}{f(\varphi)\,d\varphi}$
  integrals.}\BibitemShut {Stop}%
\bibitem [{com({\natexlab{a}})}]{comment_on_Kubo}%
  \BibitemOpen
  \href@noop {} {} \bibinfo {note} {We do not consider terms
  in Kubo formulas that are provided by products of only retarded or only
  advanced Green's functions. Contributions from such terms are accumulated at
  zero momentum. Together with similar contributions from retarded-advanced (or
  advanced-retarded) correlators, they should either vanish or be included in
  renormalization of the parameters of the model (for large values of the
  parameter $\varepsilon_0\tau$). For example, a more accurate computation of
  the GD tensor that takes into account contributions from $p=0$ includes the
  following self-consistent renormalization of the exchange splitting (at
  $\alpha_{\text{\tiny R}}=0$):
  $\Delta^\prime_{\text{sd}}=\Delta_{\text{sd}}-(4\pi\tau\Delta^\prime_{\text{sd}})^{-1}\Bigl[\log{\left(\varepsilon+\Delta^\prime_{\text{sd}}\right)}-\log{\left(\varepsilon-\Delta^\prime_{\text{sd}}\right)}\Bigr]$.
  Fermi sea contributions to STT, GD, and ESR are disregarded throughout the
  paper as well. For the present model, they vanish when
  $\varepsilon>\Delta_{\text{sd}}$.}\BibitemShut {Stop}%
\bibitem [{com({\natexlab{b}})}]{comment_on_derivative}%
  \BibitemOpen
  \href@noop {} {} \bibinfo {note} {In fact, $\partial_{\bb
  v}$ is proportional to a directional derivative, with a prefactor equal to
  $\vert\bb v_{\text{d}}\vert^{-1}$.}\BibitemShut {Stop}%
\bibitem [{las()}]{last_resort_comment}%
  \BibitemOpen
  \href@noop {} {}\bibinfo {note} {As it follows from
  Eq.~(\ref{alpha_beta_gamma_def}) and the relation $\delta
  S_{\text{eff}}=-\xi_0 S$.}\BibitemShut {Stop}%
\bibitem [{com({\natexlab{c}})}]{comment_on_singularities}%
  \BibitemOpen
  \href@noop {} {} \bibinfo {note} {The conditions $w_k\in
  C_\pm$ can only be fulfilled for some particular values of $\varepsilon$.
  Since $\delta \bb S$ is a continuous function of~$\varepsilon$, one may just
  ignore such values.}\BibitemShut {Stop}%
\end{thebibliography}

\end{document}